\documentclass[preprint,showpacs,amsmath,amssymb,pra,supercriptaddress,floatfix]{revtex4}
\usepackage{graphicx}
\usepackage{dcolumn}
\usepackage{bm}

\begin{document}

\title{Multi-Channel Atomic Scattering and Confinement-Induced Resonances in Waveguides}
\date{\today}
\author{Shahpoor Saeidian}
\email[]{s_saeid@physi.uni-heidelberg.de}
\affiliation{Physikalisches Institut, Universit\"at Heidelberg, Philosophenweg 12,
69120 Heidelberg, Germany}
\author{Vladimir S.Melezhik}
\email[]{melezhik@thsun1.jinr.ru} \affiliation{Bogoliubov Laboratory of Theoretical Physics, Joint Institute for Nuclear Research, Dubna, Moscow Region 141980, Russian Federation}%
\author{Peter Schmelcher}
\email[]{Peter.Schmelcher@pci.uni-heidelberg.de}
\affiliation{Physikalisches Institut, Universit\"at Heidelberg,
Philosophenweg 12, 69120 Heidelberg, Germany}%
\affiliation{Theoretische Chemie, Institut f\"ur Physikalische Chemie,
Universit\"at Heidelberg, INF 229, 69120 Heidelberg, Germany}%

\date{\today}
\begin{abstract}\label{txt:abstract}
We develop a grid method for multi-channel scattering of atoms
in a waveguide with harmonic confinement. This 
approach is employed to extensively analyze the transverse excitations
and deexcitations as well as resonant scattering processes.
Collisions of identical bosonic and fermionic as well as distinguishable
atoms in harmonic traps with a single frequency $\omega$
permitting the center-of-mass (c.m.) separation are explored in depth.
In the zero-energy limit and single mode regime
we reproduce the well-known confinement-induced resonances (CIRs) for bosonic, fermionic and heteronuclear collisions.
In case of the multi-mode regime up to four open transverse channels are considered. 
Previously obtained analytical results are extended significantly here.
Series of Feshbach resonances in the transmission behaviour are identified and analyzed.
The behaviour of the transmission with varying energy and scattering lengths is discussed
in detail. The dual CIR leading to a complete quantum suppression of atomic scattering 
is revealed in multi-channel scattering processes.
Possible applications include, e.g., cold and ultracold atom-atom collisions
in atomic waveguides and electron-impurity scattering in quantum wires.
\end{abstract}

\maketitle

\section{Introduction}

During the last years, the field of ultracold few-body confined systems has progressed
remarkably. By employing optical dipole traps \cite{traps} and atom chips \cite{chips1,chips2,chips3} it is possible
to fabricate mesoscopic structures in which the atoms are freezed to occupy a single or a few lowest quantum
states of a confining potential such that in one or more dimensions the characteristic length possesses the order of the
atomic deBroglie wavelength.
These configurations can be well described by effective one-dimensional systems. Well-known examples are quantum
wires and atom waveguides or quasi two-dimensional systems such as
2D electronic gases. The quantum dynamics of such systems is strongly
influenced by the geometry of the confinement.

To control the coherent propagation of particle beams through 1D or 2D traps it is crucial to very well
understand the impact of the confinement on the collisional properties (see refs.
\cite{bolda,stock1,yurovsky1,stock2,yurovsky2,naidon,bhongale,yurovsky3,Olshanii0, 
Olshanii1,Mora1a,Mora1b,Mora2,Granger,Kinoshita,Paredes,Guenter,Kim1,Egger,Kim2,Kim3,Mel2007,Lupu1,Lupu2,Olshanii2} 
and refs.therein). Free-space scattering theory is no
longer valid in such systems and a new theory is needed.  This stimulated the development of quantum scattering
theory in low dimensions.
For a sufficiently dilute gas under a strong transverse confinement,
one may expect both the chemical potential and the thermal energy $k_BT$ being less than
the transverse level spacing.
With this assumption, the dynamics of ultracold atoms in low dimensional structures e.g. tight wave
guides has been studied using the simplification
that the atoms occupy only the ground state of the transverse confining potential.  Nevertheless, the virtual
transverse excitations in the course of the
collision process can play a crucial role in the scattering leading to the so-called
confinement-induced resonances (CIRs), predicted by Olshanii \cite{Olshanii0,Olshanii1}. S-wave
zero-energy scattering of bosons is mapped to an effective longitudinal zero-range 
pseudopotential $g_{1D}\delta(z)$ approximating the 1D atom-atom interaction in a transverse harmonic confinement.
It was shown\cite{Olshanii0,Olshanii1} that CIRs appear if the binding energy of the pseudopotential, approximating
the two-atom molecular state in the presence of the confinement, coincides with the energy spacing between
the levels of the transverse harmonic potential. In the vicinity of the CIR the coupling constant $g_{1D}$ can be tuned from
$-\infty$ to $+\infty$ by varying the strength of the confining potential over a small range.  This can result in
a total atom-atom reflection, thereby creating a gas of impenetrable bosons.
CIRs have also been studied for the three-body \cite{Mora1a,Mora1b} and the four-body \cite{Mora2}
scattering under confining potential, as well as for a pure p-wave scattering of fermions \cite{Granger}.  Experimental
evidence for the CIRs for bosons \cite{Kinoshita,Paredes} and fermions \cite{Guenter} has recently been reported. A general
analytical treatment of low-energy scattering under action of a general cylindrical confinement involving all partial
waves and their coupling, was first provided in ref.\cite{Kim1} for a spherically symmetric short-range potential.
The effect of the c.m. motion on the s-wave collision of two distinguishable atoms (i.e. a two-species mixture)
in a harmonic confinement as well as for two identical atoms in a non-parabolic confining potential has been
investigated in ref.\cite{Egger} in the zero-energy limit neglecting
the s and p wave mixing. A detailed study including the effect of the c.m. nonseparability and taking into account
the s and p wave mixing for harmonic confinement was performed in the single-mode regime
in ref. \cite{Mel2007}. Recently a so-called dual-CIR was discovered \cite{Kim2,Kim3,Mel2007}, 
which is characterized by a complete transmission (suppression of quantum
scattering) in the waveguide due to destructive interference of s and p waves although the corresponding collisions
in free space involve strong interactions.

The problem of atomic pair collisions under the action of a harmonic trap in the multi-mode regime when
the energy of the atoms exceeds the level spacing of the transverse trapping potential is much more
intricate than the single-mode regime due to several open transverse channels. It demands the
development of a multi-channel scattering theory accounting for the possible transitions between the levels of the
confining potential. Using as a starting point the formalism for scattering in restricted geometries 
suggested in refs. \cite{Lupu1,Lupu2}, the multi-channel scattering problem for bosons in a harmonic confining potential has been
analyzed analytically by Olshanii el al. \cite{Olshanii2} in the s-wave pseudopotential approximation and the
zero-energy limit. Without detailization of the interatomic interaction but using only two input parameters - the
s-wave two-body scattering length in free space and the trap frequency - they have derived an approximate
formula for the scattering amplitude describing two boson collisions confined by transverse harmonic trap in the
multi-mode regime. 

In the present work we develop a general grid method for multi-channel scattering
of identical as well as distinguishable atoms confined by a transverse harmonic trap.
The method applies to arbitrary atomic interactions, permitting a
rich spectral structure where several different partial waves are participating in the scattering
process or even the case of an anisotropy of the interaction. The only limitation is that
we consider harmonic traps with a single frequency for every atom causing a separation
of the c.m. and relative motion. With our approach we analyze transverse
excitations/deexcitations in the course of the collisional process (distinguishable or identical atoms) including all important
partial waves and their couplings due to the broken spherical symmetry. Special attention is paid to the
analysis of the CIRs in the multi-mode regimes for non-zero collision energies, i.e., to suggest a non-trivial extension of
the CIRs theory developed so far only for the single-mode regime and zero-energy limit.

In detail we proceed as follows.  Section II contains the derivation of the Hamiltonian, definitions of the
interatomic interaction and the scattering asymptotics in the confined
geometry, and a discussion of the scattering parameters and the transition probabilities characterizing the two-body
collisions in the trap.  Our computational method is outlined in Sec. III, important technical details are given in the
Appendix.  In section IV our results are presented and analyzed. A summary and conclusions are given in Sec. V.

\section{Hamiltonian and Two-body Scattering Problem in a Waveguide}\label{txt:sec2}

Let us consider collisions of two atoms under the action of the
transverse harmonic confinement. We address both cases, distinguishable and indistinguishable atoms
under the action of the same confining potential, i.e. the
trap is characterized by a single frequency $\omega$ for every atom. The
corresponding Hamiltonian is given by
\begin{equation}
H=-\frac{\hbar ^2}{2m_1}\nabla_1^2 -\frac{\hbar
^2}{2m_2}\nabla_2^2 + \frac{1}{2}m_1\omega^2\rho_1^2 +
\frac{1}{2}m_2\omega^2\rho_2^2 +V(\mathbf{r}_1-\mathbf{r}_2),
\end{equation}
where $m_i$ is the mass of the $i-{th}$ atom,
$V(\mathbf{r}_1-\mathbf{r}_2)$ is the two-body potential
describing the interaction between two colliding atoms in
free space and $\mathbf{r}_i = (\rho_i,z_i)=(r_i,\theta_i,\phi_i)$
are the coordinates of the $i-{th}$ atom.  The Hamiltonian is
separable with respect to the relative $\mathbf{r}=\mathbf{r}_1 -
\mathbf{r}_2$ and c.m.
$\mathbf{R}=(m_1\mathbf{r}_1+m_2\mathbf{r}_2)/(m_1+m_2)$
variables
\begin{equation}
H= H_{\textit{CM}} + H_{\textit{Rel}}\,\,\,,
\end{equation}
where
\begin{equation}
H_{\textit{CM}}=-\frac{\hbar ^2}{2M}\nabla_R^2 + \frac{1}{2}M\omega^2\rho_R^2,
\end{equation}
and
\begin{equation}
H_{\textit{Rel}}=-\frac{\hbar ^2}{2\mu}\nabla_r^2 +
\frac{1}{2}\mu\omega^2\rho^2 + V(r)\,\,.
\end{equation}
Here $M=m_1 + m_2$ and $\mu=m_1m_2/(m_1+m_2)$ are the total and
reduced masses respectively. The problem, thus, reduces to scattering
of a single effective particle with the reduced mass $\mu$ and collision energy
$\epsilon > \hbar \omega $, off a scatterer $V(r)$ at the origin,
under transverse harmonic confinement with frequency $\omega$
\begin{equation}\label{poth}
\left [-\frac{\hbar ^2}{2\mu}\nabla_r^2 +
\frac{1}{2}\mu\omega^2\rho^2 + V(r)\right ] \psi (\mathbf{r}) =
\epsilon\psi(\mathbf{r})\,\,.
\end{equation}
Here the energy of the relative two-body motion $\epsilon =\epsilon_{\perp}+\epsilon_{\|}$ is a sum of the
transverse $\epsilon_{\perp}$ and longitudinal collision
$\epsilon_{\|}$ energies. Due to our definition of the confining
potential,  the transverse excitation energies $\epsilon_{\perp}$ 
can take the possible values $\epsilon_{\perp}= \epsilon-\epsilon_{\|}=\hbar\omega (2n + |m| +1) > 0$
of the discrete spectrum of the 2D oscillator $\frac{1}{2}\mu\omega^2\rho^2$.

For the two-body interaction we choose a screened Coulomb potential
\begin{equation}\label{potential}
V(r)=V_0\frac{r_0}{r}e^{-r/r_0},
\end{equation}
already employed in ref.\cite{Kim2,Kim3,Mel2007} for analyzing the ultracold scattering in cylindrical waveguides
in the single-mode regime as
$\epsilon_{\perp}= \epsilon-\epsilon_{\|}$. The chosen potential (\ref{potential}) depends on two parameters - the depth $V_0<0$
and the screening length $r_0>0$. Following the computational scheme already developed in ref.\cite{Mel2007}, we
implement different spectral structures of the atomic interaction by varying the single parameter
$V_0$ for a fixed length scale $r_0$. Obvious advantages of the screened Coulomb
potential (\ref{potential}) compared to the s-wave pseudopotential used in ref. \cite{Olshanii2} that is devoted to the multi-channel
scattering of bosons are the following. First, by varying $V_0$ one
can vary the number of bound states of s-wave character in the interaction potential, and second, one can create new bound and
resonant states of higher partial wave character. As a consequence we can consider not only bosonic but also fermionic as well as
mixed collisions in a trap and including the case of higher energies. This
will, as we shall see below, permit us to investigate new regimes
and effects of multi-channel confined scattering.

Performing the scale transformation
\begin{eqnarray}\label{transformation}
r\rightarrow\frac{r}{a_0}
\quad\mathrm{,}\quad\epsilon\rightarrow\frac{\epsilon}{\epsilon_0}\quad\mathrm{,}\quad
V_0\rightarrow\frac{V_0}{\epsilon_0}\quad\mathrm{and}\quad \omega\rightarrow\frac{\omega}{\omega_0}
\end{eqnarray}
with the units $a_0 = \hbar^2/\mu V_0r_0$, $\epsilon_0 = \hbar^2/\mu a_0^2$, and
$\omega_0=\epsilon_0/\hbar$,
it is convenient to rewrite the equation (\ref{poth}) in the rescaled
form
\begin{equation}\label{rescaled scheroedinger equation }
\left [-\frac{1}{2}\nabla_r^2 + \frac{1}{2}\omega^2\rho^2 + V(r)\right ]\psi (\mathbf{r}) =
\epsilon\psi(\mathbf{r})\,\,,
\end{equation}
where $V$ is now the correspondingly scaled potential and we fix $\mu=1$ and $r_0=1$ in the
subsequent consideration. In the asymptotic region, $|z|\rightarrow\infty$, where the transverse trapping potential dominates the interaction
potential, the axial and transverse motions decouple and the asymptotic wavefunction can be written as a
product of the longitudinal and transverse $\phi_{n,m}(\rho,\varphi)$
components with
\begin{equation}\label{transvers Hamil. eigenfunction}
\phi_{n,m}(\rho ,\varphi)=[\frac{\pi a^2_{\bot}(n+|m|)!}{n!}]^{-1/2}(\frac{\rho}{a_{\bot}})^{|m|}e^{-\rho^2/(2a^2_{\bot})}
L^{|m|}_{n}(\rho^2/a^2_{\bot})e^{im\varphi}
\end{equation}
being the eigenfunctions of the transverse trapping Hamiltonian
\begin{equation}\label{Ham2D}
H_{\bot}=-\frac{1}{2}(\frac{\partial^2}{\partial\rho^2}+\frac{1}{\rho}\frac{\partial}{\partial\rho}+\frac{1}{\rho^2}\frac{\partial^2}{\partial\varphi^2}) + \frac{1}{2}\omega^2\rho^2,
\end{equation}
with the corresponding eigenvalues $\epsilon_{\bot}=\omega (2n+|m|+1)$ and the angular momentum projection $m$ onto the $z$-axis.
Here $L^m_n(x)$ are the generalized Laguerre polynomials and $a_{\bot}=1/\sqrt{\omega}$, is the transverse oscillator length. The radial and azimuthal quantum numbers $n$ and $m$
of the 2D harmonic oscillator independently take the values $n = 0, 1, 2, ... \infty \quad\mathrm{and}\quad m=0, \pm1, \pm2, ...
\pm\infty$. Due to the axial symmetry of the transverse confinement and
the spherical symmetry of the interatomic interaction (\ref{potential})
the angular momentum component along the $z$-axis is conserved. Therefore, the problem (\ref{rescaled scheroedinger equation })
is reducible to a 2D one by separating the $\varphi$-variable and can be solved for every $m$
independently. The quantum number $n$ is a good one only in the asymptotic region
$|z|\rightarrow\infty$ and used for the definition of the initial (incident) asymptotic
state i.e. channel
\begin{eqnarray}\label{initial asymptotic sate}
\psi^{in}_{n,m}(\mathbf{r})= e^{ik_{n}z}\phi_{n,m}(\rho ,\varphi)\,\,,
\end{eqnarray}
of two infinitely separated atoms confined in a transverse state $<\rho\varphi|nm>$. In addition to the quantum
numbers $n$ and $m$ the asymptotic scattering state is defined also by the momentum $k_n$ of the channel
\begin{equation}\label{wave vector k}
k_{n}=\frac{2}{a_{\bot}}\sqrt{\varepsilon - n -\frac{|m|}{2}}\,\,,
\end{equation}
which we express through the dimensionless energy
$\varepsilon=\epsilon/(2\omega)-1/2$. It is clear that the integer
part of the dimensionless energy $\varepsilon$ coincides with the
number $n_{e}$ of open excited transverse channels. For $n\leq n_{e}$ we have $(\varepsilon - n-\frac{|m|}{2})>0$.
The spherically symmetric interatomic interaction (\ref{potential}) mixes 
different transversal channels and leads to transitions $n\rightarrow n'$ between the
open channels $n,n'\leq n_{e}$, i.e. to transverse
excitation/deexcitation processes during the collisions.

Assuming the system to be initially in the channel $n$, the asymptotic wavefunction takes at $\mid z\mid \rightarrow +\infty$ the
form \cite{Olshanii2}
\begin{eqnarray}\label{full asymptotic wave}
\psi_{n,m}(\mathbf{r})=e^{ik_{n}z}\phi_{n,m}(\rho ,\varphi)+\sum_{n'=0}^{n_{e}}\left [f^e_{nn'} + sgn(z)f^o_{nn'}\right
]e^{ik_{n'}|z|}\phi_{n',m}(\rho ,\varphi),
\end{eqnarray}
where $f^e_{nn'}$ and $f^o_{nn'}$ are the matrix elements of the inelastic scattering
amplitudes for the even and odd partial waves, respectively, which describe transitions between the channels $n$ and
$n'$. For a bosonic (fermionic) collision just the symmetric (antisymmetric) part of (\ref{full asymptotic wave}) should
be considered.

It is clear that the scattering amplitude depends also on the index $m$ which, however, remains unchanged
during the collision due to the axial symmetry
of the problem. Hereafter we consider only the case $m = 0$ and the index is omitted in the following.

Using the asymptotic wavefunction (\ref{full asymptotic wave}) and the
total current conservation one obtains
\begin{equation}\label{current conservation}
\sum^{n_{e}}_{n'=0}\left ( T_{nn'} + R_{nn'} -\delta_{nn'} \right ) = 0
\end{equation}
for the inelastic transmission (reflection) coefficients $T_{nn'}$ ($R_{nn'}$). $k_n$ ($k_{n'}$) is the  initial (final) relative wave vector and $n$ ($n'$) the transverse
excitation numbers according to eq.(\ref{wave vector k}). We have \cite{Olshanii2}
\begin{equation}\label{transmission}
T_{nn'} = \Theta [\varepsilon - n']\frac{k_{n'}}{k_{n}}|\delta_{n,n'}+f^e_{nn'}+f^o_{nn'}|^2
\end{equation}
\begin{equation}\label{reflection}
R_{nn'} = \Theta [\varepsilon -
n']\frac{k_{n'}}{k_{n}}|f^e_{nn'}+f^o_{nn'}|^2\,\,,
\end{equation}
where $\Theta(x)$ is the Heavyside step-function.  The transition probability $W_{nn'}$, characterizing the transverse
excitation/deexcitation, into a particular channel $n'$ from the initial state $n$ is given by the sum of the corresponding transmission and reflection
coefficients
\begin{equation}\label{transition probability}
W_{nn'}=T_{nn'}+R_{nn'}.
\end{equation}
 Due to the time reversal symmetry of the Hamiltonian we have $T_{nn'}=T_{n'n}$, $R_{nn'}=R_{n'n}$ and
$W_{nn'}=W_{n'n}$. The total transmission (reflection) coefficient $T= \sum_{n^{\prime}} T_{nn^{\prime}}$ ($R= \sum_{n^{\prime}} R_{nn^{\prime}}$) is given by the sum of the transmission (reflection)
coefficients of all the open channels.  Eq.(\ref{current conservation}) leads to $T+R=1$.

\section{Numerical approach}\label{txt:sec4}

To obtain the observable quantities $T_{nn'}\,,\,R_{nn'}$ and
$W_{nn'}$ of the scattering process, we have to calculate the
matrix elements $f_{nn'}$ of the scattering amplitude $\hat{f}$ by
matching the numerical solution of the Schr\"{o}dinger equation
(\ref{rescaled scheroedinger equation }) with the scattering
asymptotics (\ref{full asymptotic wave}). To integrate this multi-channel scattering
problem in two dimensions $r$ and $\theta$ ($z$ and $\rho$)
we adopt the discrete-variable method suggested in
ref.\cite{Mel91} for solving nonseparable 2D scattering
problems. This approach was applied in ref.\cite{Mel2003} to the case of a 3D
anisotropic scattering problem of ultracold atoms in external laser fields.

First, we discretize the 2D Schr\"{o}dinger equation
(\ref{rescaled scheroedinger equation }) on a 2D grid of 
angular $\{\theta_j\}_{j=1}^{N_\theta}$ and radial $\{r_j\}_{j=1}^N$
variables. The angular grid points $\theta_j$ are defined as
the zeroes of the Legendre polynomial
$P_{N_{\theta}}(\cos\theta)$ of the order $N_{\theta}$. Using the completeness
property of the normalized Legendre polynomials which remains valid also on the chosen angular grid
\begin{equation}\label{completeness}
\sum^{N_{\theta} -1}_{l=0}P_l(\cos \theta_j)P_l(\cos
\theta_{j'})\sqrt{\lambda_j\lambda_{j'}}=\delta_{jj'}\,\,,
\end{equation}
where $\lambda_j$ are the weights of the Gauss quadrature, we
expand the solution of equation (\ref{rescaled scheroedinger equation }) in the basis
$f_j(\theta)=\sum^{N_{\theta} -1}_{l=0}P_l(\cos \theta)(\mathbf{P}^{-1})_{lj}$ according to
\begin{equation}\label{grid expansion of wave function}
\psi(r, \theta) = \frac{1}{r}\sum ^{N_{\theta
}}_{j=1} f_j(\theta) u_j(r)\,\,.
\end{equation}
Here $\mathbf{P}^{-1}$ is the inverse of the $N_{\theta}\times
N_{\theta}$ matrix $\mathbf{P}$ with the matrix elements defined
as $\mathbf{P}_{jl}=\sqrt{\lambda_j}P_l(\cos \theta_j)$. Due to
this definition one can use the completeness relation
(\ref{completeness}) in order to determine the matrix elements
$(\mathbf{P^{-1}})_{lj}$ as
$(\mathbf{P^{-1}})_{lj}=\sqrt{\lambda_j}P_l(\cos
\theta_j)$. It is clear from (\ref{grid expansion of wave
function}) that the unknown coefficients $u_j(r)$ in the
expansion are the values $\psi(r,\theta_j)$ of the 
two-dimensional wave function $\psi(r,\theta)$ at the grid points
$\theta_j$ multiplied by $\sqrt{\lambda_j} r$. Near the origin $r\rightarrow 0$ we have $u_j(r)\simeq
r \rightarrow 0$ due to the definition (\ref{grid expansion of
wave function}) and the demand for the probability distribution
$\mid\psi(r,\theta_j)\mid^2$ to be bounded. Substituting
(\ref{grid expansion of wave function}) into (\ref{rescaled
scheroedinger equation }) results a system of $N_{\theta}$
Schr\"{o}dinger-like coupled equations with respect to
the $N_{\theta}$-dimensional unknown vector
$\mathbf{u}(r)=\{\lambda^{1/2}_ju_j(r)\}^{N_{\theta}}_1$
\begin{equation}\label{Sheroedinger-like equation}
[\mathbf{H}^{(0)}(r)+2(\epsilon\mathbf{I}-\mathbf{V}(r))]\mathbf{u}(r)=0,
\end{equation}
where
\begin{equation}\label{H^0}
\mathbf{H}^{(0)}_{jj'}(r)=\frac{d^2}{dr^2}\delta_{jj'}-\frac{1}{r^2}\sum^{N_{\theta}-1}_{l=0}
\mathbf{P}_{jl}l(l+1)(\mathbf{P}^{-1})_{lj'}\,\,, \end{equation}
\begin{equation}\label{V-matrix elements}
\mathbf{V}_{jj'}(r)=V(r,\theta_j)\delta_{jj'}=\{V(r)+
\frac{1}{2}\omega^2\rho^2_j\}\delta_{jj'},\quad \rho_j=r\sin \theta_j\,\,,
\end{equation}
and $\mathbf{I}$ is the unit matrix.
We solve the system of equations (\ref{Sheroedinger-like equation}) on the quasi-uniform radial
grid \cite{Mel97}
\begin{equation}\label{r-mapping}
r_j = R\frac{e^{\gamma x_j}-1}{e^{\gamma}-1}\,\,,\,\,j=1,2,...,N
\end{equation}
of $N$ grid points $\{r_j \}$ defined by mapping $r_j\in(0,R\rightarrow
+\infty]$ onto the uniform grid $x_j\in (0,1]$ with the equidistant
distribution $x_j - x_{j-1} = 1/N$. By varying
$N$ and the parameter $\gamma > 0$ one can choose more adequate
distributions of the grid points for specific interatomic and
confining potentials.

By mapping the initial variable $r$ in Eq.(\ref{Sheroedinger-like equation}) onto $x$ we obtain
\begin{equation}\label{Sheroedinger-like equation2}
[\mathbb{H}^{(0)}(x)+2\{\epsilon\mathbf{I}-\mathbf{V}(r(x))\}]\mathbf{u}(r(x))=0\,\,,
\end{equation}
with
\begin{equation}\label{H^02}
\mathbb{H}^{(0)}_{jj'}(x)=f^2(x)\delta_{jj'}\left (\frac{d^2}{dx^2}-\gamma\frac{d}{dx} \right )
-\frac{1}{r^2(x)}\sum^{N_{\theta}-1}_{l=0}\mathbf{P}_{jl}l(l+1)(\mathbf{P}^{-1})_{lj'}\,\,, \end{equation}
where
\begin{equation}
f(x) = \frac{e^{\gamma} - 1}{Re^{\gamma x}\gamma}\,\,.
\end{equation}
The uniform grid with respect to $x$ gives  6-order accuracy for applying a
7-point finite-difference approximation of the derivatives in the
equation (\ref{Sheroedinger-like equation2}) . Thus, after the
finite-difference approximation the initial 2D Schr\"{o}dinger
equation (\ref{rescaled scheroedinger equation }) is reduced to
the system of N algebraic matrix equations
\begin{eqnarray}\label{LU}
\sum_{p=1}^3\mathbb{A}^j_{j-p}\mathbf{u}_{j-p}+
[\mathbb{A}^j_j+2\{\epsilon\mathbf{I}-\mathbf{V}_j\}]\mathbf{u}_j+\sum_{p=1}^3\mathbb{A}^j_{j+p}\mathbf{u}_{j+p}=0
\,\,,\,\, j=1,2,...,N-3\nonumber\\
\mathbf{u}_j+\alpha_j^{(1)}\mathbf{u}_{j-1}+\alpha_j^{(2)}\mathbf{u}_{j-2}+
\alpha_j^{(3)}\mathbf{u}_{j-3}+\alpha_j^{(4)}\mathbf{u}_{j-4}=\mathbf{g}_j\quad\quad j=N-2, N-1, N
\end{eqnarray}
where each coefficient $\mathbb{A}^j_{j'}$ is a
$N_{\theta}\times N_{\theta}$ matrix, each $\alpha_j$ is a diagonal $N_{\theta}\times N_{\theta}$ matrix and each 
$\mathbf{g}_j$ is a $N_\theta$-dimensional vector. Here the functions
$\mathbf{u}_{-3}$, $\mathbf{u}_{-2}$, $\mathbf{u}_{-1}$ and
$\mathbf{u}_{0}$ in the first three equations of the system (for
$j=1,2$ and 3) are eliminated by using the ``left-side'' boundary
conditions: $\mathbf{u}_{0}=0$ and $\mathbf{u}_{-j}=\mathbf{u}_{j}$
($j=1,2,3$). The last three equations in this system for $j=N,N-1$ and $N-2$ are the ``right-side'' boundary
conditions approximating at the edge points $r_{N-2},r_{N-1}$ and $r_N=R$ of the radial grid, the
scattering asymptotics (\ref{full asymptotic wave}) for the desired wave function
$\mathbf{u}(r_j)$. In order to construct the ``right-side'' boundary
conditions (\ref{LU}) at $j=N-2,N-1$ and $N$ we used an idea
of ref.\cite{Mel2003} i.e. the asymptotic behaviour
(\ref{full asymptotic wave}) at the edge points $r_{N-2},r_{N-1}$ and $r_N=R$ are considered as
a system of vector equations with respect to the unknown vector
$f_{nn'}$ of the scattering amplitude for a fixed $n$. By
eliminating the unknowns $f_{nn'}$ from this system we implement the
``right-side'' boundary conditions defined by Eqs.(\ref{LU}) at $j=N-2,N-1$ and
$N$ (see Appendix A).

The reduction of the 2D multi-channel scattering problem to the
finite-difference boundary value problem (\ref{LU}) permits one to
apply efficient computational methods. Here we use, in the spirit
of the $LU$-decomposition \cite{Press}, and the sweep method \cite{Gelfand} (or the Thomas algorithm \cite{fia}), a fast implicit
matrix algorithm which is briefly described in Appendix B. The
block-diagonal structure of the matrix of the coefficients in the system
of equations(\ref{LU}) with the width of the diagonal band equal
to $7\times N_\theta$ makes this computational scheme an efficient one.

Solving the problem (\ref{LU}) for the defined initial vector $k_n$ and
a fixed $n$ from the possible set $0\leq n\leq n_{e}$  we first calculate the vector function
$\psi(k_n,r,\theta_j)$. Then, by  matching the calculated vector $\psi(k_n,R,\theta_j)$ with the asymptotic behaviour
(\ref{full asymptotic wave}) at $r=R$, we calculate the $n$-th row of the scattering amplitude matrix
$f_{nn'}$ describing all possible transitions $n\rightarrow
n'=0,1,...,n_{e}$. This procedure is repeated for the next
$n$ from $0\leq n\leq n_{e}$. After calculating all the elements $f_{nn'}$ of the
scattering amplitude we obtain any desired scattering parameter $T$,$R$ or $W$ .

\section{Results and Discussion}\label{txt:sec5}

With the above-described method being implemented we have analyzed the two-body scattering under the
transverse harmonic confinement for both cases of identical and
distinguishable colliding atoms. For confined scattering of
identical atoms one has to distinguish the bosonic and fermionic cases. In the case of two colliding bosons
the two-body wave function must be symmetric and only even scattering
amplitude provides us with a nonzero contribution. First, we show that our
result for the special case $\varepsilon < 1$ of a single-channel
scattering is in agreement at $\varepsilon\rightarrow 0$ with the
s-wave pseudopotential  approach \cite{Olshanii0}, and,
particularly, reproduces s-wave CIR predicted and
analyzed in refs. \cite{Olshanii0,Olshanii1,Olshanii2,Kim2,Mel2007} for bosons. Then we extend
our consideration to the multi-channel scattering $\varepsilon > 1$. We demonstrate
that our results are in a good agreement in the limit of a long-wavelength trap
$\omega=2\pi c/\lambda \rightarrow 0$ with the analytical
expression given in \cite{Olshanii2} which has been obtained
in the s-wave pseudopotential approach for the zero-energy limit.
The range of validity of the analytical investigation in ref. \cite{Olshanii2} is explored. Next
we present results for multi-channel scattering of two
fermions under transverse harmonic confinement. For a
fermionic collision the two-body wave function is antisymmetric, i.e. only the
odd scattering amplitude is nonzero. In the special case of
single-channel scattering we reproduce the p-wave CIR for fermions
\cite{Granger}. These results are also in agreement with our
previous investigations in refs.\cite{Kim2,Kim3,Mel2007} performed within a
wave-packet propagation method \cite{Mel2007}.  Finally we
consider the confined multi-channel scattering of two
distinguishable atoms. In this case both even and odd
amplitudes contribute to the scattering process.
\begin{figure*}[h]
$\begin{array}{c}
\includegraphics[height=6cm,width=8cm]{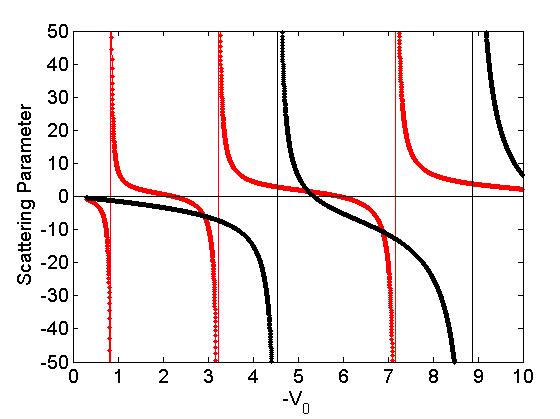}\\
\end{array}$
\caption{s- and p-wave scattering parameters $a_s$ (red) and
$V_p$ (black) as a function of the potential depth $V_0$ for
the free-space effective potential $V(r)+l(l+1)/(2 r^2)$. 
Divergences correspond to the appearance of new bound (s-wave
scattering) or shape resonant (p-wave scattering) states in the
effective potential. All quantities are given in the units
(\ref{transformation}).} \label{fig1}
\end{figure*}

For modeling different interatomic interactions in the subsequent
sections we vary the depth $V_0$ of the potential
(\ref{potential}) in the wide range $-10<V_0<0$ for a fixed width
$r_0=1$. In free space, this potential being
superimposed with the centrifugal term $l(l+1)/(2 r^2)$ makes an
effective potential which may support some \textit{even} or
\textit{odd} bound states depending on the value of the quantum
number $l$ and the parameter $V_0$. For $l\neq 0$ there might be
also some \textit{shape resonances} for certain relative energies.
These \textit{shape resonances} may enhance strongly the
contribution of $l\neq 0$ partial waves in the energy domain where
one would have expected a pure $l=0$ scattering.

It is known that in the zero-energy limit, when the scattering process
does not depend on the details of the potential, the collision can
be described by a single parameter: the s-wave scattering length
$a_s=-\lim_{k\rightarrow 0}\tan\delta_s(k)/k$ for $l=0$ (bosonic
collision) and the p-wave scattering volume $V_p=-\lim_{k\rightarrow
0}\tan\delta_p(k)/k^3$ for $l=1$ (fermionic collision). For sufficiently
low collision energy the contribution of the partial waves with
larger $l$ can be neglected.  In Fig.\ref{fig1} we have plotted
$a_s$ and $V_p$ in the region $-10<V_0<0$. 
Fig.\ref{fig1} demonstrates the rich spectral structure of the
chosen form of the interatomic interaction: the scattering
parameters $a_s$ and $V_p$ can be positive or negative and they
diverge for the values of $V_0$ corresponding to the appearance of
new bound states. In the case of p-wave scattering the increase of
the depth $V_0$ of the potential first leads to a shape
resonance which approaches zero energy and finally transforms
to a p-wave bound state.

\subsection{Multi-channel scattering of bosons}

\begin{figure*}[h]
$\begin{array}{cc}
\includegraphics[height=6cm,width=8cm]{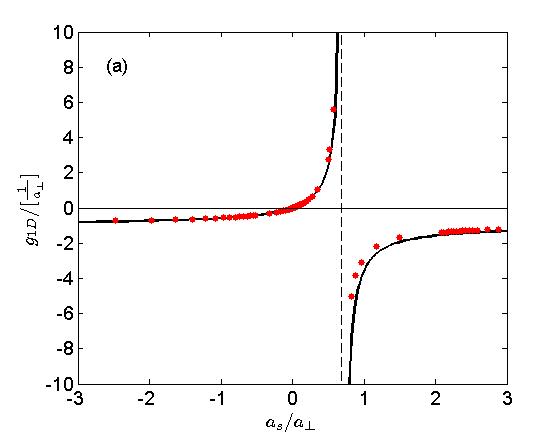} & \includegraphics[height=6cm,width=8cm]{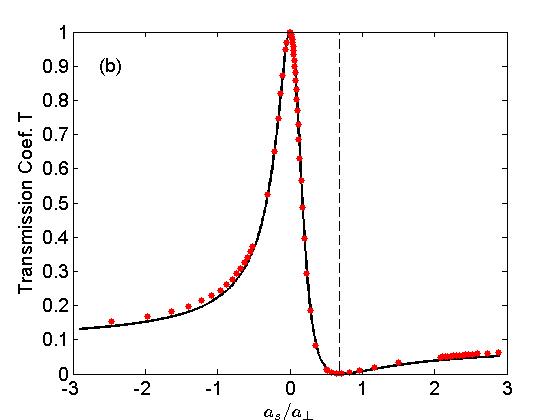}\\
\includegraphics[height=6cm,width=8cm]{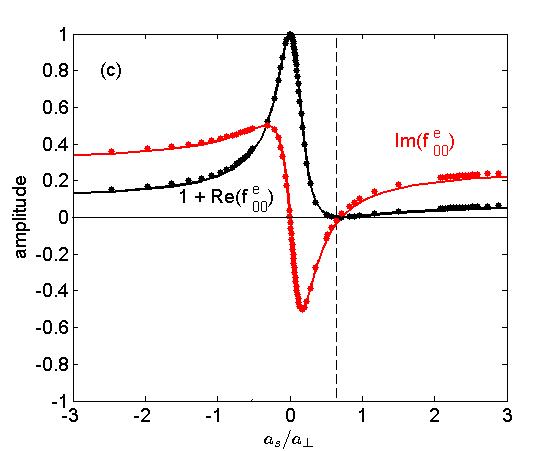} \\
\end{array}$
\caption{(a) The effective quasi-1D coupling constant $g_{1D}$,
(b) the transmission coefficient $T$ and (c) the scattering
amplitude $f^e_{00}$ as a function of the scattering length
$a_s$ for one-channel scattering of two bosons under a harmonic
confining potential with $\omega=0.002$ for a longitudinal relative
energy $\epsilon_{\|}=0.0002$ together with the analytical results
obtained for the s-wave pseudo-potential zero-energy limit(solid curves
)\cite{Olshanii0,Olshanii1,Olshanii2}. The constant $V_0$ is varied
in the region $-2.30< V_0 <-0.32$.}\label{fig2}
\end{figure*}

For two bosons colliding in a transverse harmonic
confinement the scattering wave function is symmetric with respect
to the exchange $z\rightarrow -z$, i.e.  $f^o_{nn'}=0$ in Eq.(\ref{full
asymptotic wave}).  We consider multi-channel scattering with
the dimensionless energy
$\varepsilon=\epsilon/(2\omega)-1/2<4$ (\ref{wave vector
k}) permitting the collisional transverse excitations/deexcitations
$n\rightarrow n'<4$ up to four open channels (four-mode
regime). For comparison with analytical results
\cite{Olshanii1,Olshanii2} obtained in the s-wave pseudopotential approach, we have
extracted the effective quasi-1D coupling constant $g_{1D}=\lim_{k\rightarrow
0}Re\{f^e_{00}(k)\}/Im\{f^e_{00}(k)\}k/\mu$ as
well as the transmission coefficient $T=T_{00}$(\ref{transmission}) and the scattering
amplitude $f^e_{00}$(\ref{full asymptotic wave}), as a function of the scattering length $a_s$
in the single-mode regime ($0 < \varepsilon < 1$). The calculated
parameters are presented along with the analytical results in
Fig.\ref{fig2} for $\omega = 0.002$ and the longitudinal relative
energy $\epsilon_{\|}=\epsilon - \epsilon_{\bot}=0.0002$.
Fig.\ref{fig2}(a) shows the coupling constant $g_{1D}$ as a
function of the scattering length $a_s$. Our numerical result
clearly exhibit a singularity at $a_s/a_{\bot}\approx 1/C$ with
$C=-\zeta (1/2)=1.4603..$, which corresponds to the well-known
s-wave CIR \cite{Olshanii0,Olshanii1}. In Fig.\ref{fig2}(b) we
present the transmission coefficient $T$ versus $a_s$.  The
transmission coefficient $T$ goes to unity (total transmission)
when $a_s$ tends to zero (i.e. no interaction between the
atoms), while at the CIR position, it exhibits the well-known
minimum (blocking of the atomic current by the CIR).
Fig.\ref{fig2}(c) shows the scattering amplitude $f^e_{00}$ as a
function of $a_s$. The amplitude approaches zero at
$a_s/a_{\bot}=0$ and $-1$ at the CIR position , which results in
total transmission and total reflection respectively.  In general
the presented values of $f^e_{00}$ are in very good
agreement with the analytical results.

\begin{figure*}[h]
$\begin{array}{cc}
\includegraphics[height=6cm,width=8cm]{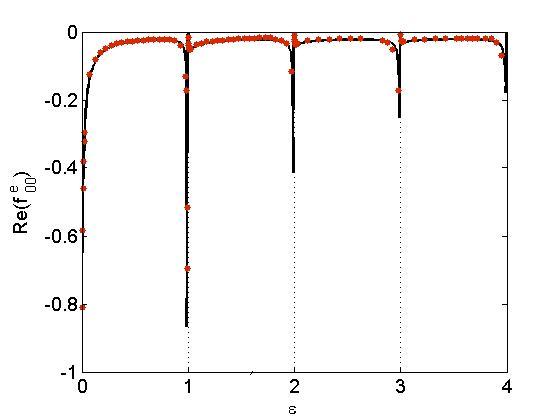}&\includegraphics[height=6cm,width=8cm]{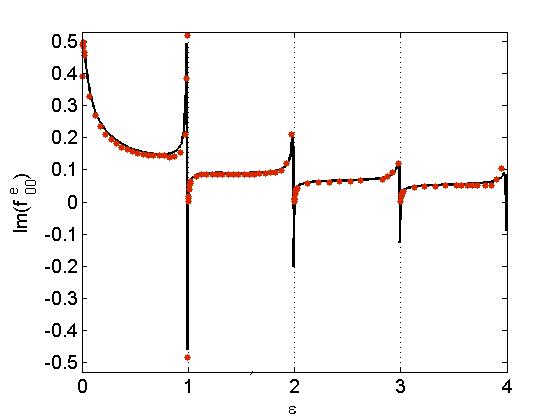}\\
\end{array}$
\caption{The scattering amplitude $f^e_{00}$ (dots)
as a function of the dimensionless energy $\varepsilon$ along with
the analytical results (solid curves) for $\omega =0.0002$ and
$V_0=-3.0$ ($a_s = -8.95$).} \label{fig3}
\end{figure*}

\begin{figure*}[h]
$\begin{array}{cc}
\includegraphics[height=6cm,width=8cm]{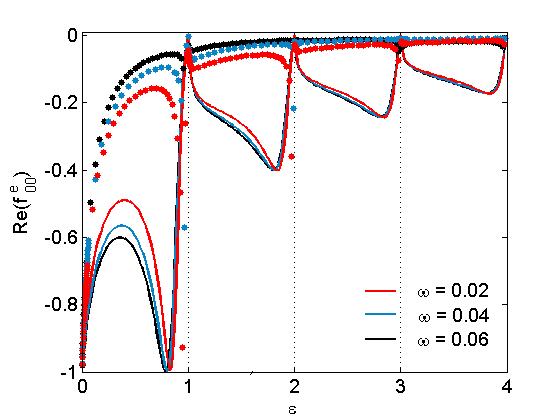} \includegraphics[height=6cm,width=8cm]{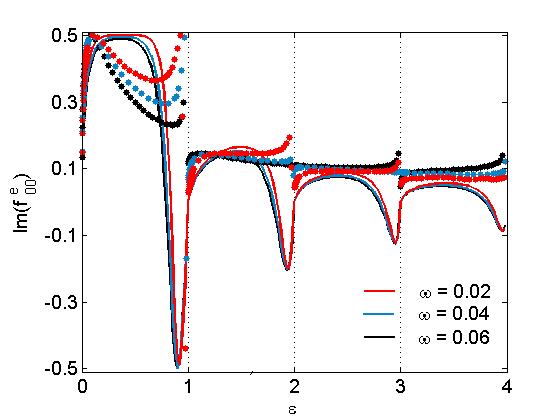}\\
\end{array}$
\caption{The scattering amplitude $f^e_{00}$ (dots)
as a function of the dimensionless energy $\varepsilon$ along with
the analytical results (solid curves) for $\omega =0.02$ (red),
$\omega =0.04$ (blue) and $\omega =0.06$ (black).  The amplitudes
have been calculated for $V_0=-3.0$ ($a_s= -8.95$).}
\label{fig4}
\end{figure*}

In the multi-mode regime our results are in good agreement
with the analytical ones  obtained within the s-wave pseudopotential approach in the zero-energy limit\cite{Olshanii2}
for long-wavelength traps (tight confinement: $\omega=2\pi c/\lambda \rightarrow 0$).
In Fig.\ref{fig3} we present our scattering amplitude
$f^e_{00}$ along with the analytical results as a function of the
dimensionless energy $\varepsilon$ for $\omega=0.0002$ and
$V_0=-3.0$. Apart from energies close to the channel thresholds, the real and imaginary part
of the scattering amplitude $f^e_{00}$ show a monotonous behaviour: $Re(f^e_{00})$ is monotonically
increasing and approaching zero asymptotically whereas $Im(f^e_{00})$ decays monotonically and also
approaches zero for large values of $\varepsilon$. The peak structure located at integer
values of $\varepsilon$ is due to the resonant scattering once a new previously closed channel
opens with increasing energy. There is a good agreement between our results and the
analytical ones given by Eq.(6.9) in ref. \cite{Olshanii2} for the complete range
$0 < \varepsilon < 4$. However, we encounter major deviations
with increasing $\omega$ except for narrow regions close to the channel thresholds
(as $\varepsilon$ approaches integer values)
for the real parts of the scattering amplitudes (see Fig.\ref{fig4}). These deviations are
most presumably due to the energy dependence of the s-wave scattering length which is neglected
in ref.\cite{Olshanii2}.


\begin{figure*}[h]
$\begin{array}{cc}
\includegraphics[height=6cm,width=8cm]{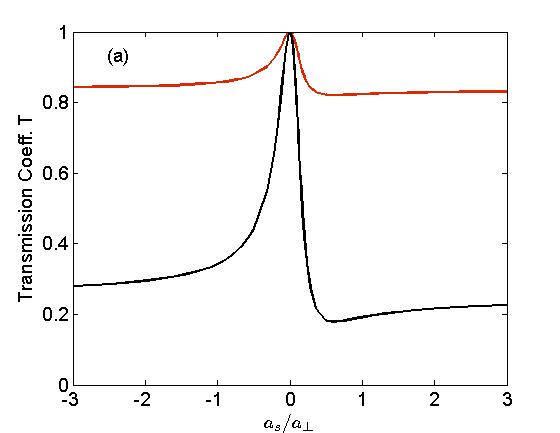} \includegraphics[height=6cm,width=8cm]{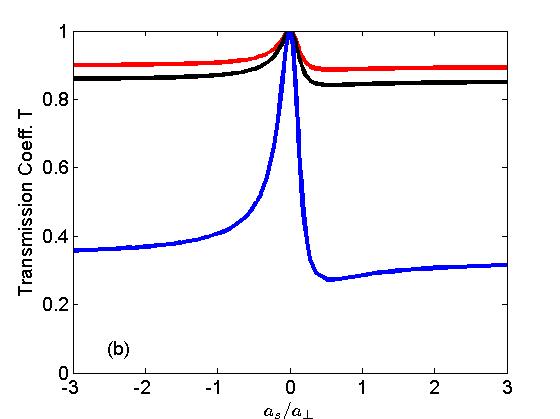}\\
\end{array}$
\caption{The total transmission coefficient $T$ for a bosonic
collision in a harmonic confinement $\omega = 0.002$ as a
function of $a_s/a_{\perp}$ for (a) two-open channels with $\varepsilon=1.05$
being initially in the transverse ground state $n=0$ (red) and first excited
state $n=1$ (black), and (b) three-open channels with $\varepsilon=2.05$
being initially in the ground state $n=0$ (red), first excited
state $n=1$ (black) and second excited state $n=2$ (blue). }
\label{fig5}
\end{figure*}

For the multi-channel regime ($\varepsilon > 1$) we find a
strong dependence of the total transmission coefficient $T$ on the
population of the initial state $n$, except for the case $a_s/a_{\bot}\rightarrow 0$ of
noninteracting bosons in free space. Fig.\ref{fig5} shows the
calculated transmission coefficients(\ref{transmission}) as a function of $a_s/a_{\perp}$ for
$\omega =0.002$ for the (a) two-mode regime with $\varepsilon =1.05$ and
(b) the three-mode regime with $\varepsilon =2.05$. Note, that all these cases correspond to near-threshold
collision energies if the maximal integer is subracted from $\varepsilon$. Similar to the
single-mode regime, $T$ goes to unity (total transmission) when
$a_s$ tends to zero.  We also encounter a
minimum.  However, the value of the transmission at the minimum is not zero anymore in the multi-mode regime, the larger the number of open channels, the position of the minima will be more shifted to the left. For a
fixed value of the ratio $a_s/a_{\perp}\neq 0$, a lower initially
populated transverse level $n$, leads to a larger total transmission.

\begin{figure*}[h]
$\begin{array}{cc}
\includegraphics[height=6cm,width=8cm]{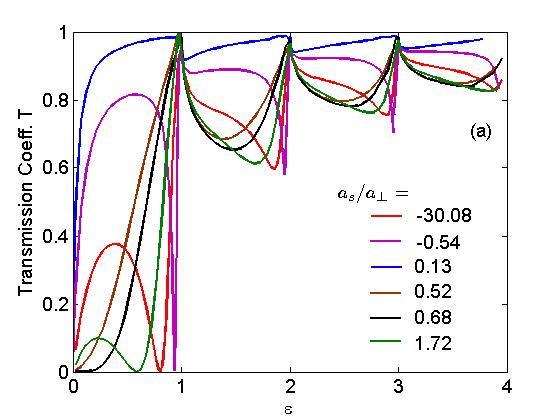}\includegraphics[height=6cm,width=8cm]{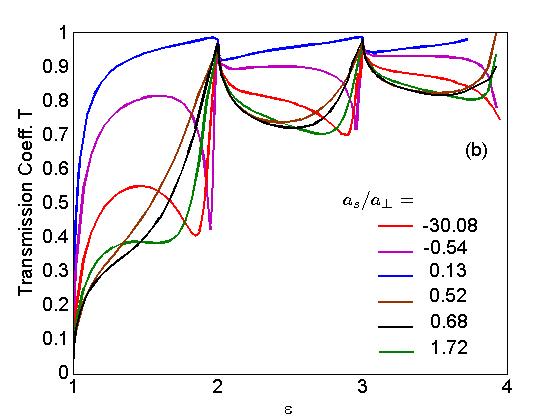}\\
\end{array}$
\caption{The total transmission coefficients $T$ for bosonic collisions as a function of the
dimensionless energy $\varepsilon$ for the two cases of the system being initially in the ground $n=0$ (a)
and first excited $n=1$ (b) transverse states, for several ratios of $a_s/a_{\bot}$ and
$\omega = 0.002$. The black curve corresponds to $a_s/a_{\bot}=1/C$ for which 
the zero-energy CIR in the single-mode regime is encountered.} \label{fig6}
\end{figure*}

In Fig.\ref{fig6} we present the transmission 
as a function of the dimensionless energy $\varepsilon$ for the
two cases of the system being initially in the ground $n=0$ (a) or first
excited $n=1$ (b) transverse states, for different values of the ratio 
$a_s/a_{\bot}$.  With increasing collision energy the transmission coefficient exhibits
a nonmonotonic behaviour: we observe a sequence of minima and peaks.
For $0<\varepsilon<1$ the value of $T$ at the minimum is zero. As the number of open channels increases with
increasing collision energy $\epsilon$, the transmission values at the corresponding minima
increase strongly. The corresponding transmission peaks $T=1$ are located 
at the channel thresholds. The shape of the transmission 'valleys' in between two
integer values of $\varepsilon$ as well as the positions of the minima strongly depend on
the ratio $a_s/a_{\bot}$ which can be changed by varying the strength of the interatomic interaction
$V_0$ or the trap frequency $\omega$. Both, changing $V_0$ and/or $\omega$
leads to energetical shifts of the bound states (in particular for the excited transversal
channels) of the atoms in the presence of the confinement.
If the collision energy coincides with a bound state of the corresponding closed
channel we encounter an occupation of the closed channel in the course of the scattering
process, i.e. a Feshbach-resonance occurs.
The interpretation of the minimum of the transmission $T$ in terms of a Feshbach-resonance at the
point $a_s/a_{\bot}=1/C$ for the zero-energy limit was provided in refs.
\cite{Olshanii0,Olshanii1}, where it was shown that the origin of the CIR is
an intermediate occupation of a bound state belonging to an excited transverse (closed) channel.

\begin{figure*}[h]
\includegraphics[height=6cm,width=8cm]{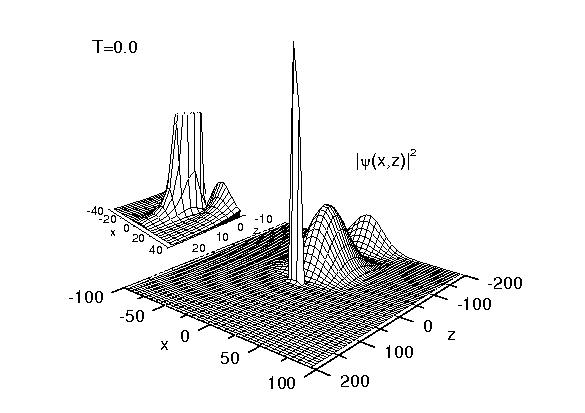}
\caption{The probability density $|\psi(x,z)|^2$ for bosonic collisions as a function of $x$ and $z$
at $a_s/a_{\bot}=0.68$ for $\varepsilon=0.05$ - zero-energy CIR . The corresponding transmission
values are also indicated. The result has been obtained for $\omega=0.002$.} \label{fig6E}
\end{figure*}

To demonstrate that the above-discussed behaviour (minima) of the transmission coefficient $T(\varepsilon)$ in certain regions of
$\varepsilon$ is due to Feshbach resonances we have analyzed the 
probability density of the scattering wave function of the atoms in the trap. In the single-mode regime and in the zero-energy
limit we encounter the well-known CIR: Fig.\ref{fig6E} shows the corresponding
probability density $|\psi(x,z)|^2$ for an initial transverse ground state $n=0$ and $\varepsilon=0.05$ as
well as $a_s/a_{\bot}=0.68$.
For small $|z|$ one observes additional two pronounced peaks along the transverse ($x$-)
direction corresponding to the occupation of the bound state (with the binding energy $\varepsilon_{n=1}^B\sim 0$) in the first excited closed channel of the transverse
potential. The probability density tends to zero as $z\rightarrow +\infty$. This leads to a zero of the transmission
$T(\varepsilon)$ for $\varepsilon\rightarrow 0$ (see Fig.\ref{fig6}(a)) corresponding to the zero-energy CIR.

\begin{figure*}[h]
$\begin{array}{cc}
\includegraphics[height=6cm,width=8cm]{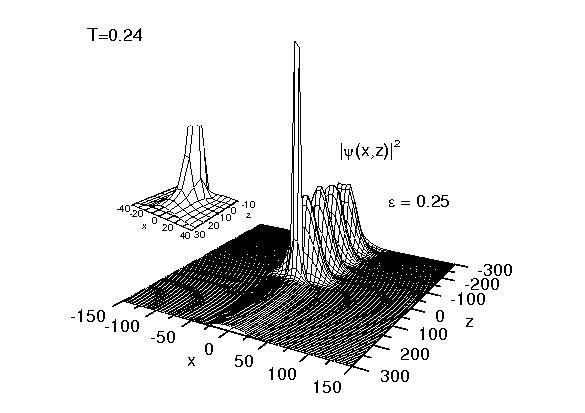}\includegraphics[height=6cm,width=8cm]{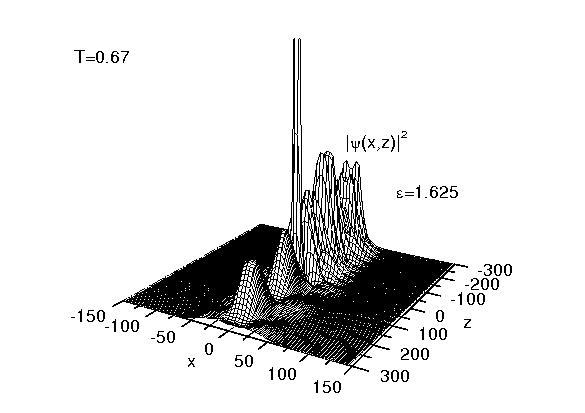}\\
\includegraphics[height=6cm,width=8cm]{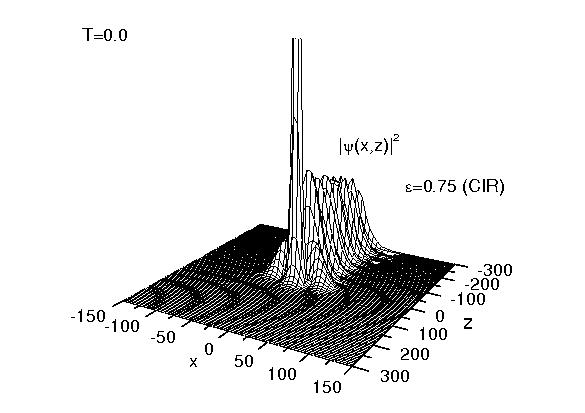}\includegraphics[height=6cm,width=8cm]{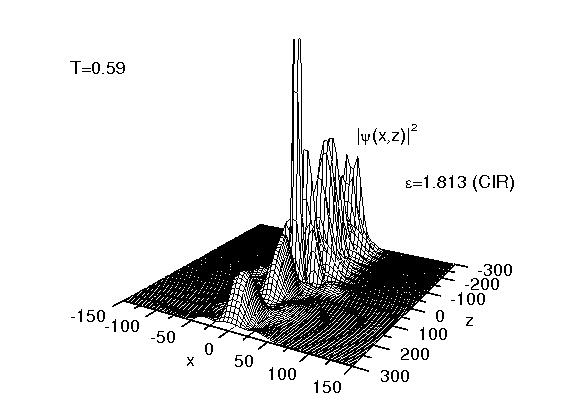}\\
\includegraphics[height=6cm,width=8cm]{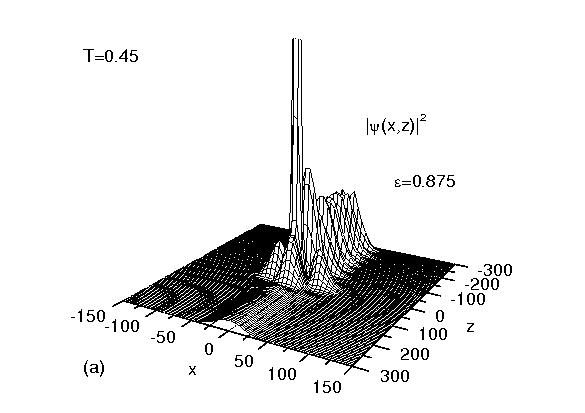}\includegraphics[height=6cm,width=8cm]{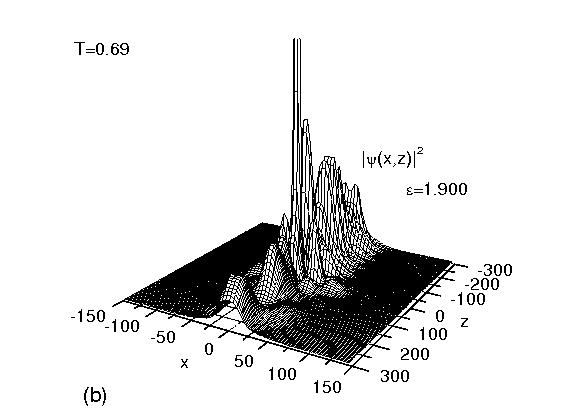}\\
\end{array}$
\caption{The probability density $|\psi(x,z)|^2$ for bosonic collisions as a function of $x$ and $z$
at $a_s/a_{\bot}=+4.39$ for two cases of
the single-mode regime (a) and two-mode regime (b) with different values of $\varepsilon$. The corresponding transmission
values are also indicated. All subfigures are for $\omega=0.002$ and $n=0$.} \label{figE6}
\end{figure*}

In Fig.\ref{figE6} we show the probability densities $|\psi(x,z)|^2$ at $a_s/a_{\bot}=+4.39$ 
for several values of the dimensionless energy $\varepsilon$ for (a) the single- and (b) two-mode regimes for collisions with initial transverse state $n=0$.  The corresponding probability density exhibits for small values of $|z|$ additional two (for the single-mode regime) and four (for the two-mode regime) pronounced peaks with respect to the transverse ($x$-)direction as $\varepsilon$ approaches the CIR-position.  This demonstrates the occupation of bound states of higher,
namely first excited (with binding energy $\varepsilon_{n=1}^B$) and second excited (with binding energy $\varepsilon_{n=2}^B$) channels in the course of the scattering process.
The corresponding transmission values are also indicated in Fig.\ref{figE6}. Zero transmission is observed also
for $a_s/a_{\bot}=+4.39$ and $\varepsilon = 0.75$ with no probability density being present for large positive
values of $z$, see Fig.\ref{figE6}(a).

The energy of the bound state $\varepsilon_{n=1}^B$, that leads to a minimal transmission due to a resonant scattering process, 
changes with varying $a_s/a_{\bot}$ as follows. It is below $\varepsilon =0$ for $0<a_s/a_{\bot}<0.68$ and
consequently no minimum is encountered for $T(\varepsilon)$ in the range $0< \varepsilon <1$ (see fig.
\ref{fig6}). At the position of the zero-energy CIR it is located just above the threshold $\varepsilon = 0$
leading to a zero transmission for $\varepsilon \rightarrow 0$. For $a_s/a_{\perp} > 0.68$ the bound state
energy is somewhere in between the channel thresholds $\varepsilon = 0$ and $\varepsilon =1$ whereas for $a_s/a_{\perp} < 0$
it is below but close to $\varepsilon =1$ leading again to a corresponding minimum of $T$.
For both cases, $a_s/a_{\perp} < 0$ and $a_s/a_{\perp} > 0.68$, an increase of $a_s/a_{\perp}$ leads to 
a narrow transmission well and the corresponding transmission minimum is shifted towards the next higher
channel threshold.  The dependence of $\varepsilon_{n=1}^B$ on the parameter $a_s/a_{\bot}$ is in agreement with the pseudopotential analysis given in ref.\cite{Olshanii1} (see Fig. 2 of this ref.). 

\begin{figure*}[h]
$\begin{array}{cc}
\includegraphics[height=6cm,width=8cm]{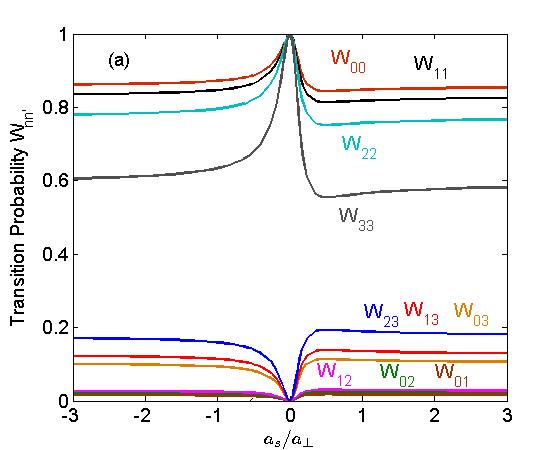}\includegraphics[height=6cm,width=8cm]{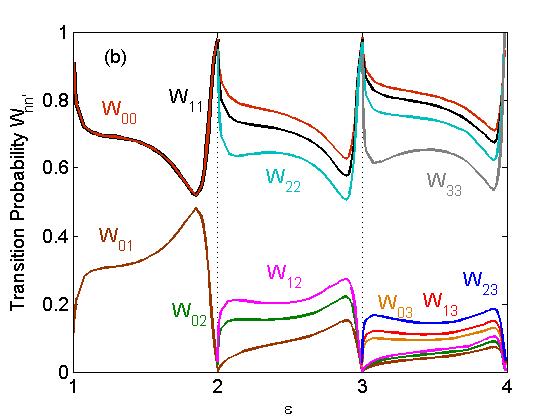}\\
\end{array}$
\caption{ (a) The transition probabilities  $W_{nn'}$ as a function of $a_s/a_{\perp}$ in the four mode regime for
$\varepsilon = 3.05$.  (b) The calculated transition probabilities  $W_{nn'}$ as a function of $\varepsilon$ up to
four open channels for $a_s/a_{\bot}=-30.08$. $\omega =0.002$ for both subfigures.} \label{fig7}
\end{figure*}

In Fig.\ref{fig7}(a) we show the transition probabilities $W_{nn'}$ as a function of $a_s/a_{\perp}$ for $\omega =0.002$
and $\varepsilon = 3.05$  corresponding to four open channels.  We observe that the probability of remaining at the same
initial state, $W_{nn}$ (i.e. elastic scattering) is in the complete range of the ratio $a_s/a_{\bot}$ much larger than
the probability of a transition into a different state, $W_{nn'}$
(i.e. inelastic scattering).  With increasing $n$ or $n'$ the inelastic transition probabilities
$W_{nn'}$ increase but the elastic probabilities $W_{nn}$ decrease. In an inelastic (elastic) collision
$W_{nn'}$ ($W_{nn}$) goes to zero (unity) as $a_s$ tends to zero. $W_{nn'}$ ($W_{nn}$) possess
a maximum (minimum) at the resonance position $a_s/a_{\perp} \approx 0.35$ consistent with the
minimum of $T(\varepsilon)$ at $\varepsilon = 3.05$. 
It is instructive to see how the distribution of the initial flux among the open channels changes due to pair collisions
as a function of the collision energy.  Fig.\ref{fig7}(b) shows the transition probabilities $W_{nn'}$ as a function of
the dimensionless energy up to four open channels.  The probability of elastic scattering remains larger than that
of inelastic scattering in the complete range of the energy. For two open channels the elastic collision probability
$W_{nn}$ is independent of the initial state ($W_{00}= W_{11}$). For a higher number of open channels
$W_{nn}$ is decreasing with increasing initial value of $n$. Near the thresholds, the probabilities of the 
inelastic (elastic) transitions $W_{nn'}$ ($W_{nn}$) go to zero (unity).

\subsection{Multi-channel scattering of fermions}

\begin{figure*}[h]
$\begin{array}{c}
\includegraphics[height=6cm,width=8cm]{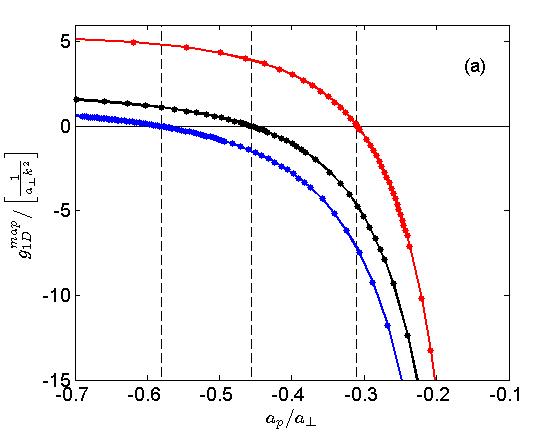}\\ 
\includegraphics[height=6cm,width=8cm]{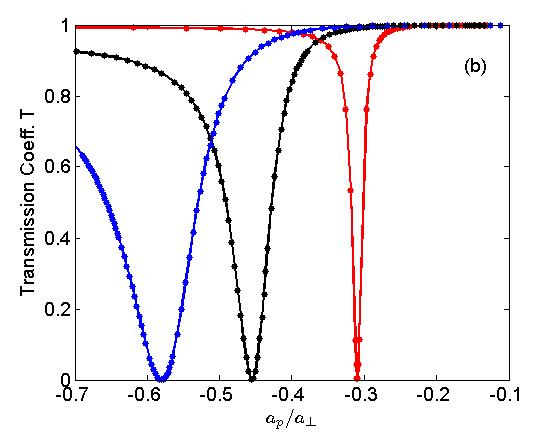}\\ 
\includegraphics[height=6cm,width=8cm]{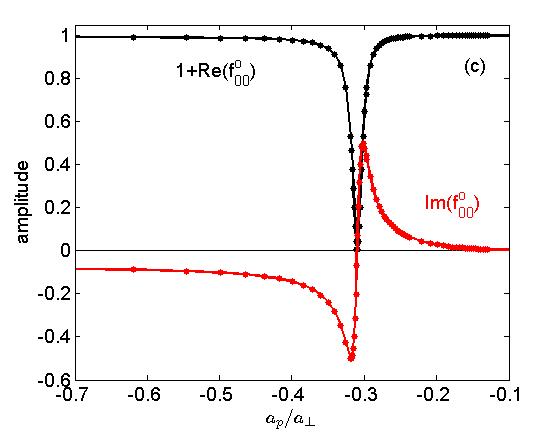}\\
\end{array}$
\caption{ (a-b) The \textit{mapped coupling constant} $g^{map}_{1D}$ and the transmission coefficient
$T$ as a function of the scattering length $a_p$ for single-channel scattering of two fermions under a harmonic
confinement $\omega =0.002$ and for the longitudinal energy $\epsilon_{\parallel} =0.0002$ (red),
$\omega =0.02$, $\epsilon_{\parallel} =0.002$ (black) and $\omega =0.06$, $\epsilon_{\parallel} =0.002$ (blue).  (c) The scattering amplitude
$f^o_{00}$ as a function of the scattering length $a_p$ for $\omega =0.002$ and $\epsilon_{\parallel} =0.0002$.
The corresponding constant $V_0$ is varied in the region $-4.54< V_0 <-4.47$.}\label{fig8}
\end{figure*}

In this section we focus on fermionic collisions in harmonic traps.
In this case the interatomic wave function is anti-symmetric with respect to the interchange of the two fermions
and the even amplitude $f^e_{nn'}$ in eqs.(\ref{full asymptotic wave}) is zero.
We have analyzed the multi-channel scattering of fermions up to four
open transverse channels for different interatomic interactions,  by varying the potential strength $V_0$ in the
vicinity of the value $V_0=-4.54$ (see Fig.1) generating a resonant p-wave state in free space.  Fig. \ref{fig8}
shows corresponding results for the single-mode regime.  Fig. \ref{fig8}(a) shows the \textit{mapped
coupling constant} $g^{map}_{1D}=\lim_{k\rightarrow
0}Im\{f^o(k)\}/Re\{f^o(k)\}\mu k$ \cite{Granger,Mel2007} as a function of
the p-wave scattering length $a_p=\sqrt[3]{V_p}$.  The
\textit{mapped coupling constant} $g^{map}_{1D}$ goes to zero at
the position of the mapped CIR \cite{Granger}, which for $\omega = 0.002$ and
$\epsilon_{\parallel} = 0.0002$  is equal to $a_p/a_{\bot}=-0.31$.  The
position of the mapped CIR obviously depends on the values of $\omega$ and
$\varepsilon$.  In Fig.\ref{fig8}(b) we have plotted the
transmission coefficient $T$ as a function of $a_p$.  The
transmission coefficient exhibits a minimum and an accompanying well (i.e. the
blocking due to the resonance) at the position of the CIR, and tends to unity
(i.e. total transmission) for $a_p$ far from the CIR-position. For
larger $\omega$ the well becomes wider, and its minimum is shifted to the left,
see also ref.\cite{Kim2}.  Fig. \ref{fig8}(c) shows the
scattering amplitude $f^o_{00}$ as a function of $a_p$ for
$\omega=0.002$ and $\epsilon_{\parallel}=0.0002$.  The amplitude approaches zero 
far from the CIR position and $-1$ at the CIR position , which results
in total transmission, $T=|1+f^o_{00}|^2\rightarrow 1$ (i.e. the
fermions do not scatter each other) and total reflection,
$T=|1+f^o_{00}|^2\rightarrow 0$ (i.e. strongly interacting and
impenetrable fermions), respectively.

\begin{figure*}[h]
$\begin{array}{cc}
\includegraphics[height=6cm,width=8cm]{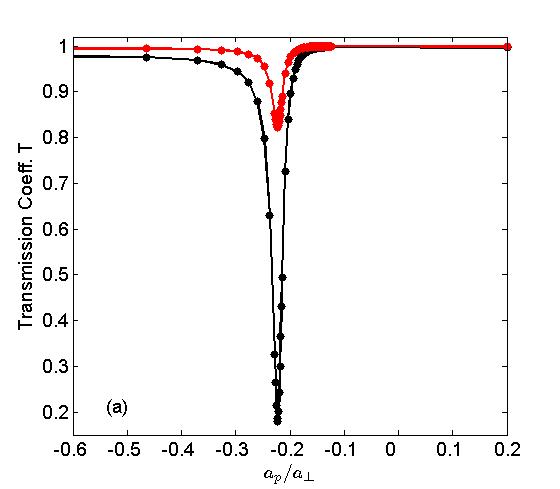}\includegraphics[height=6cm,width=8cm]{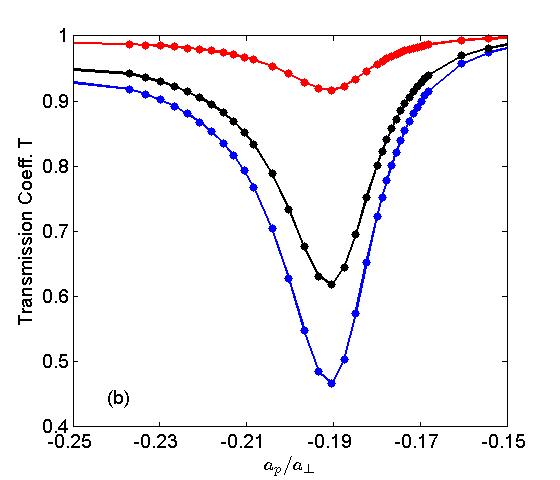}\\
\end{array}$
\caption{Total transmission coefficient $T$ in a fermionic collision as a function of $a_p$ for $\omega = 0.002$ for (a) two-open channels with $\varepsilon=1.05$ being initially in the ground state $n=0$ (black) and first excited state $n=1$ (red), and (b) three-open channel with $\varepsilon=2.05$ being initially in the ground state $n=0$ (blue), first excited state $n=1$ (black) and second excited state $n=2$ (red).} \label{fig9}
\end{figure*}

\begin{figure*}[h]
$\begin{array}{cc}
\includegraphics[height=6cm,width=8cm]{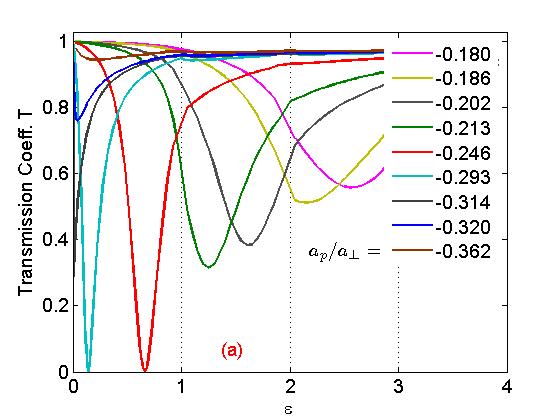}\includegraphics[height=6cm,width=8cm]{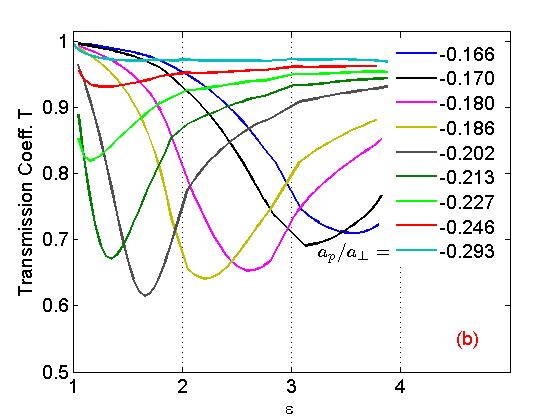}\\
\end{array}$
\caption{Total transmission coefficient $T$ for a fermionic collision as a function of the dimensionless
energy $\varepsilon$ for the two cases of the system being initially in the ground $n=0$ (a) and first excited $n=1$ (b) transverse
states, for several ratios of $a_p/a_{\bot}$. We have used $\omega = 0.002$.} \label{fig10}
\end{figure*}

In Fig.\ref{fig9} we present the total transmission coefficient as a function of $a_p$ in the two-mode regime for
$\varepsilon=1.05$ (a) and three-mode regime for $\varepsilon=2.05$ (b) for $\omega =0.002$.  Similar to the single-mode
regime, $T$ exhibits a minimum and accompanying well, however the position of the minimum is shifted to the right and
the value at the minimum is nonzero. For the lower degree of transversal excitation, we observe a deeper and narrower
transmission well.  With increasing energy the transmission well becomes wider and more shallow and its position is shifted to larger values of $a_p/a_{\bot}$. This is also demonstrated in Fig.\ref{fig10}, where
the transmission coefficient is plotted as a function of $\varepsilon$ for several values of $a_p/a_{\bot}$ for the
two cases of being initially in the ground $n=0$ (a) and the first excited $n=1$ (b) transverse states for
$\omega = 0.002$.  We observe that for any number of open channels, $T$ exhibits a minimum for some value of $a_p/a_{\bot}$.
With increasing $a_p/a_{\bot}$ the transmission well becomes more shallow (except for the single-mode regime) and wider.
In contrast to the bosonic case, there is no specific threshold behaviour.
This is a consequence of the fact that the relative motion does
not feel the interatomic interaction in the closed channels which are strongly screened by the centrifugal
repulsion playing a dominant role for near-threshold collision energies or in other words: We encounter a 
weak coupling of the different scattering channels. 


\begin{figure*}[h]
$\begin{array}{c}
\includegraphics[height=6cm,width=8cm]{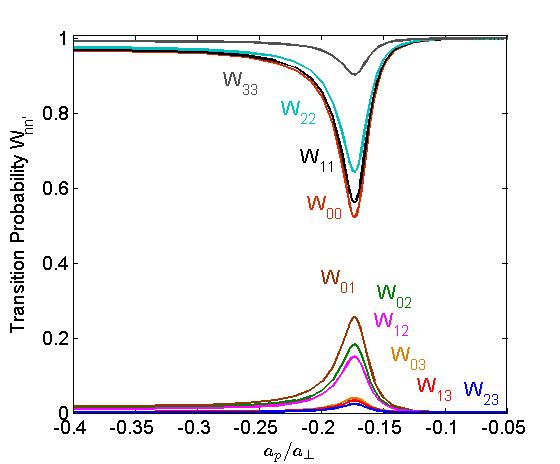}\\
\end{array}$
\caption{Transition probabilities  $W_{nn'}$ in
fermionic collisions as a function of $a_p/a_{\bot}$ for four open channels.
$\omega =0.002$ and $\epsilon = 0.0142$ are employed.}\label{fig11}
\end{figure*}

In Fig.\ref{fig11} we  show the transition probabilities
$W_{nn'}$ as a function of $a_p/a_{\bot}$ for four open channels. We see
that the probability of an elastic scattering process (i.e. to remain in
the same transversal state) is much larger than that of an
inelastic collision (i.e. the transition to a different transversal state).  For an elastic collision the probability $W_{nn}$ shows a
minimum and corresponding well which becomes wider and more shallow with increasing initial quantum number $n$ (i.e., with the population
of a higher excited initial transversal state). For an inelastic collision $W_{nn'}$ ($n \ne n'$) exhibits a
peak which becomes less pronounced as the quantum numbers $n$ or $n'$ increase.

\subsection{Multi-channel scattering of distinguishable particles}

\begin{figure*}[h]
$\begin{array}{c}
\includegraphics[height=6cm,width=8cm]{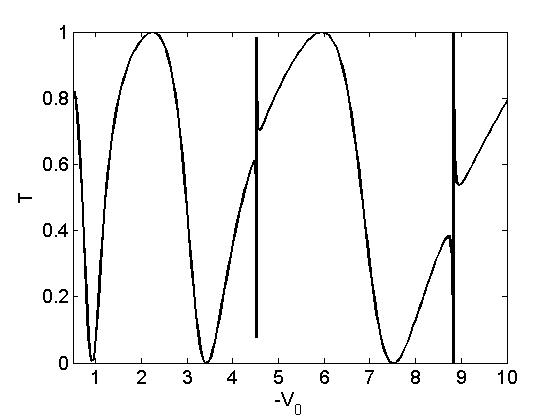}\\
\end{array}$
\caption{Transmission $T$ as a function
of the depth $-V_0$ of the potential (\ref{potential}) for two
distinguishable atomic species for $\omega = 0.002$ for
$\epsilon_{\parallel} = 0.0002$. Units according to
eq.(\ref{transformation}).}\label{fig12}
\end{figure*}

\begin{figure*}[h]
$\begin{array}{c}
\includegraphics[height=6cm,width=8cm]{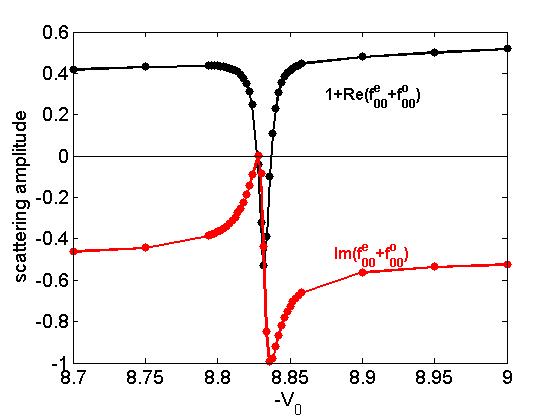}\\
\end{array}$
\caption{Scattering amplitude as a function of the
depth $-V_0$ of the potential (\ref{potential}) for two distinguishable
atomic species for $\omega = 0.002$ and $\epsilon_{\parallel} =
0.0002$. Units according to eq.(\ref{transformation}).}\label{fig13}
\end{figure*}

In this section we analyze the multi-channel scattering of two
distinguishable particles in a harmonic trap with the same trap
frequency $\omega_1 = \omega_2=\omega$ allowing the separation of
the c.m. motion.  Such a case corresponds to the different atomic species
confined by the same potential as it is the case e.g. for two different
isotopes in the same optical dipole trap. The two-body scattering
wave-function of the distinguishable atoms does not possess a well-defined
symmetry with respect to the reflection $z\rightarrow -z$, i.e.
both s- and p-state contributions ($f^e$ and $f^o$) must be taken into account
for the scattering amplitude. Fig.\ref{fig12}
shows our results for the transmission coefficient $T$ in the
single-mode regime ($0<\varepsilon <1$), which is plotted as a
function of the tuning parameter $-V_0$ of the interparticle interaction.
In general the scattering
process can not be described by a single scattering length $a_s$
or $a_p$ for this case.  In regions with a negligible p-wave
contribution (see the regions in Fig.\ref{fig1} with $V_p\rightarrow
0$), $T$ exhibits a behavior similar to bosonic
scattering. We observe the well-known s-wave CIRs which lead to zeros
of the transmission $T$ at the positions $a_s/a_{\bot} = 1/C$
and tend to unity when $a_s$ goes to zero (together with $a_p$). In
regions, where $a_s$ and $a_p$ are
comparable we observe the effect reported in ref. \cite{Kim2}:
remarkable peaks of the transmission $T=|1+f^e_{00}+f^o_{00}|^2\rightarrow 1$ i.e. almost complete
transmission in spite of the strong interatomic interaction in
free space. This is the so-called dual CIR: Quantum suppression of scattering
in the presence of confinement due to destructive interference of odd and even
scattering amplitudes. Equally minima of $T=|1+f^e_{00}+f^o_{00}|^2\rightarrow 0$ due
to the interference of even and odd scattering amplitudes under the action of the
transverse confinement can occur (see Fig.\ref{fig12}). Complete transmission corresponds to
$f^e_{00}+f^o_{00} = -1-i$ while total reflection corresponds
to $f^e_{00}+f^o_{00} = -1$. Fig.\ref{fig13} shows the corresponding
amplitudes $f^e_{00}+f^o_{00}$ as a function of $-V_0$.

\begin{figure*}[h]
$\begin{array}{c}
\includegraphics[height=6cm,width=8cm]{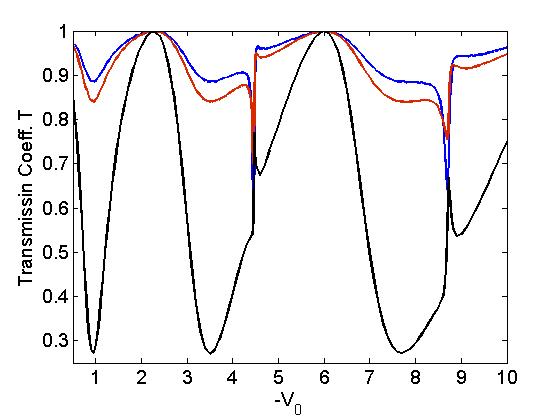}\\
\end{array}$
\caption{Transmission coefficient $T$ as a function
of the depth $-V_0$ of the potential (\ref{potential}) for two
distinguishable atomic species for $\omega = 0.002$ and $\varepsilon =
2.05$ in the three-mode regime, being initially in transverse ground
state (blue), first excited state (red) and second excited state
(black). Units according to eq.(\ref{transformation}).} \label{fig14}
\end{figure*}

Fig.\ref{fig14} shows the total transmission coefficient versus
$-V_0$ in the three-mode regime for $\omega = 0.002$ and
$\varepsilon=2.05$. Similar to the single-mode regime,  when $a_p$ is
negligible compared to $a_s$, $T$ behaves analogously to the
case of a bosonic collision, and tends to unity (complete transmission)
when $a_s$ tends to zero, while at the s-wave CIR position, it exhibits a
minimum with a nonzero value. For a lower degree of transversal excitation 
of the initial state, we encounter a larger transmission coefficient. For
the same reasons as in the single-mode regime, we observe in the regions of
$V_0$ where the p-wave scattering length $a_p$ is comparable to
$a_s$, sharp peaks of $T$. However in contrast to the
single-mode regime we do not observe complete transmission i.e.
$T\neq 1$ (see e.g. $T$ at $V_0 = -8.85$ in Fig.\ref{fig14}).

\begin{figure*}[h]
$\begin{array}{c}
\includegraphics[height=6cm,width=8cm]{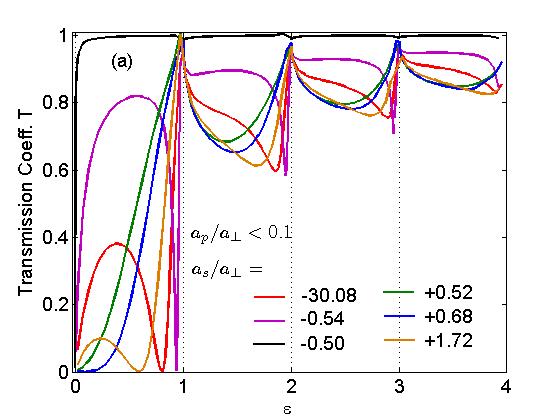}\\\includegraphics[height=6cm,width=8cm]{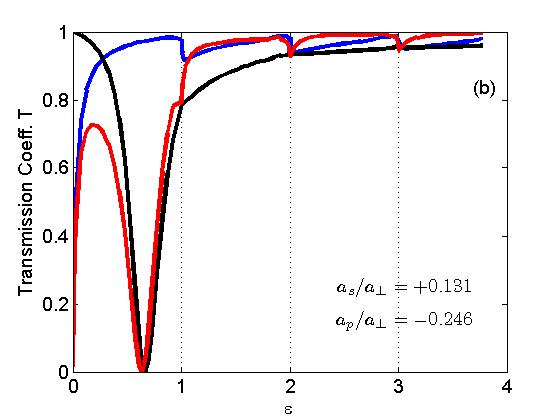}\\\includegraphics[height=6cm,width=8cm]{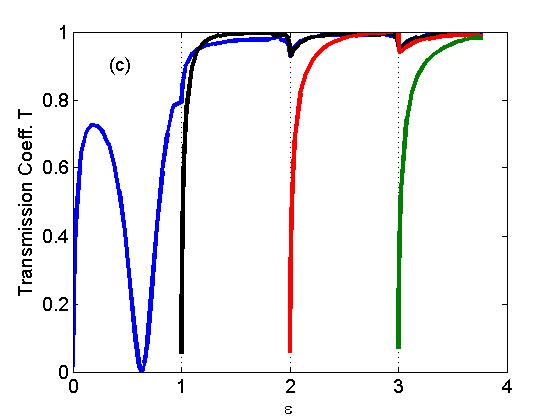}\\
\end{array}$
\caption{Transmission coefficient $T$ as a function
of the dimensionless energy $\varepsilon$ for scattering of two
distinguishable particles in a harmonic confinement 
$\omega = 0.002$ (a) for several ratios $a_s/a_{\bot}$ in the
case where $a_p$ is negligible compared to $a_s$, (b) for $-V_0 =
-4.505$ when both scattering lengths $a_s$ and $a_p$ are
comparable (here the transmission coefficients for bosonic- (blue)
and fermionic- (black) collision are also provided) and (c) for $-V_0
= -4.505$ when the system is initially in the ground transversal state (blue),
in the first excited state (black), in the second excited state
(red) and in the third excited state (green).} \label{fig15}
\end{figure*}

Fig.\ref{fig15} presents the transmission
coefficients as a function of the dimensionless energy
$\varepsilon$ for $\omega = 0.002$.  Fig.\ref{fig15}(a) shows the
results for several ratios of $a_s/a_{\bot}$ if $|a_p|$ is small
compare to $|a_s|$.  Here the system is initially in the transversal
ground state.  The behaviour of the transmission coefficient is similiar
to the case of two-boson scattering (see \ref{fig6}-a).
Fig.\ref{fig15}(b) shows the transmission for the scattering of two
distinguishable atoms together with the results for bosonic-
(blue) and fermionic- (black) collisions, for $-V_0 = -4.505$ when
both scattering length $a_s$ and $a_p$ are large.  In the
limit $\varepsilon \rightarrow 0$ the behaviour of the transmission
for the distinguishable and the bosonic case are very similar
while in the vicinity of the energy of the p-wave
shape-resonance $\varepsilon \sim 0.7$,  apart from a small
shift to larger energies, we can find a complete coincidence of the
transmission behaviour for the case of distinguishable and fermionic
scattering. At the threshold energy $\varepsilon =1$ the latter two transmission
curves cross. Fig.\ref{fig15}(c) shows the 
transmission coefficient versus $\varepsilon$ for $-V_0 = -4.505$ 
initially occupying the ground state $n=0$
(blue), the first excited state $n=1$ (black), the second excited
state $n=2$ (red) and the third excited state $n=3$ (green).
In the limit $\varepsilon_{\parallel}\rightarrow 0$, $T$ drops rapidly to zero.
Apart from the single mode regime, it is for $\varepsilon_{\parallel} >> 0$
approximately constant with $T\approx 1$.

\begin{figure*}[h]
$\begin{array}{c}
\includegraphics[height=6cm,width=8cm]{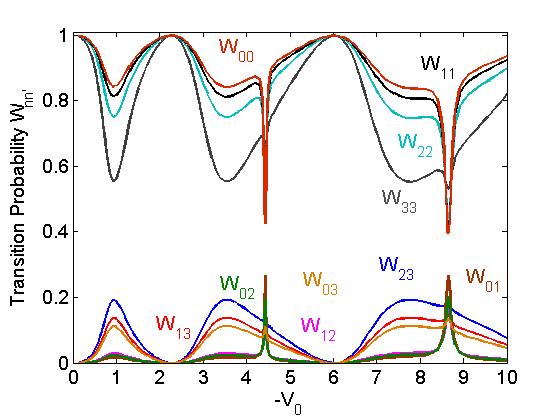}\\
\end{array}$
\caption{Transition probabilities  $W_{nn'}$ for
quasi-1D scattering of distinguishable particles in a harmonic
trap $\omega =0.002$, as a function of $-V_0$ for
four open channels and $\varepsilon = 2.05$. $V_0$ is given in units
of eq.(\ref{transformation}).} \label{fig16}
\end{figure*}

In fig.\ref{fig16} we present the transition
probabilities $W_{nn'}$ as a function of $-V_0$ for four open
channels.  For the regions of $V_0$ where $a_p$ is negligible compared to
$a_s$, a very good agreement of the transmission behaviour for distinguishable
and bosonic atoms is observed. In regions where $a_s$ and $a_p$ are comparable,
similar to the fermionic case, we observe narrow and deep wells (for elastic
collision) and narrow as well as strongly pronounced peaks (for inelastic collision).

\section{Summary and conclusions}\label{txt:sec6}

We have analyzed atomic multi-channel scattering in a
2D harmonic confinement. Identical bosonic and fermionic scattering as well as scattering
of distinguishable atoms in traps with the same frequency for the different
species have been explored. Equal frequencies allows us to separate the c.m. and
relative motion.

Firstly we reproduced the well-known s-wave CIR for bosonic collision in the single-mode
regime\cite{Olshanii0,Olshanii1,Olshanii2, Kim2,Kim3,Mel2007}.
Next bosonic collisions in the multi-channel regime including elastic and inelastic processes
i.e. transverse excitations and deexcitations have been investigated.
Transmission coefficients as well as transition probabilities for
energies covering up to four open channels are reported on. It
is shown that the transmission coefficient as a function of the
scattering length exhibits a minimum at the CIR position. 
Except for $0<a_s/a_{\bot}<1/C$ in the first channel, the transmission curves show, with varying
energy, a minimum and accompanying well.
For a single open channel the value of the minimum of the transmission is zero. 
Increasing the degree of transverse excitation leads to an increase of the transmission minima.
A distinct threshold behaviour is observed at every $\varepsilon =n$.

For $a_s/a_{\bot}<0$ the accompanying transmission well is more pronounced compared
to the case $a_s/a_{\bot}>0$ and it is also deeper in the case of
several open channels. The position of the minimum is closer to the upper
threshold again when compared to the case $a_s/a_{\bot}>0$. With increasing ratio $a_s/a_{\bot}$ for either $a_s/a_{\bot}>0$ or $a_s/a_{\bot}<0$, the corresponding transmission well becomes also more pronounced 
i.e. we encounter a deeper and narrower well for more open channels and the energetical position of the
minimum moves to the neighboring upper threshold.
Our results for the transition probabilities $W_{nn'}$ show that the probability of remaining in
the same state (elastic collision) is larger than transitions into
a different transversal state (inelastic collision).  In the case of elastic
collisions (inelastic collisions) it has to go to unity (zero) as $a_s$
tends to zero. The inelastic transition probability $W_{nn'}$ increases as the
numbers $n$ or $n'$ increase while the elastic transition probability $W_{nn'}$ decreases.

Our next focus has been the multi-channel scattering of fermions 
addressing up to four open channels in the waveguide. In the regime of a single open channel
we reproduced the well-known p-wave mapped CIR in the zero-energy limit \cite{Granger,Mel2007}.
We have analyzed the p-wave CIR dependence
on the collision energy $\epsilon$ and the trapping frequency $\omega$.
Consequently the multi-mode regime has been explored by determining
the behaviour of the transmission as well as the transition
probabilities. The dependence of the transmission on the p-wave
scattering length exhibits a minimum with zero transmission and an accompanying well
for the single-mode regime. Increasing the energy, this well
becomes more shallow (except for the single-mode regime),wider and its
position is shifted to the region of higher energies. In the
multi-mode regime the transmission well depends also on the degree of the initial transverse
excitation. Increasing the degree of initial transverse excitation
the transmission well becomes shallower and wider and the transmission is
overall increased. For a fixed number of open channels the transmission exhibits as a function of the energy 
a minimum for some value of the p-wave scattering length. Increasing $a_p/a_{\perp}$ leads to a shift
of the transmission minimum to higher energies and an increasingly shallower well. This holds except
for the single mode regime in which the minimal value is zero.

In contrast to the bosonic case, we do not observe a distinct threshold behavior. 
The transition probabilities $W_{nn'}$ for fermionic collisions show
that elastic collisions are more probable
than inelastic ones. The probability $W_{nn}$ for an elastic process shows a well which becomes wider and
shallower as the initial population of the excited states $n$ increases, while for an
inelastic collision it exhibits a peak which becomes
smaller as the channel numbers $n$ or $n'$ increase. By varying
the potential parameter $V_0$
in the viccinity of the value $V_0=-4.54$,
we found a set of shape-resonances.

Finally we analyzed the multi-channel-1D scattering of distinguishable
atoms in the trap. Here both s- and p-wave
contributions have been taken into account. We have studied the
transmission coefficient in the single-mode regime and reproduced the
CIR \cite{Olshanii0,Olshanii1,Olshanii2,Granger} as well as the
dual-CIR \cite{Kim2,Kim3,Mel2007} corresponding to total reflection
and transmission, respectively. In the multi-mode regime the transmission
versus $-V_0$ shows except for the regions where $a_p$ is comparable with
$a_s$ a behaviour analogous to the situation of bosonic scattering. We
encounter $T \rightarrow 0$ for $a_s \rightarrow 0$. 
At the position of the CIR, $T$ exhibits a minimum with a nonzero value.
The lower is the transverse excitation of the initial state, the larger is the transmission
coefficient. In regions of $V_0$ where the p-wave
scattering length $a_p$ is comparable to $a_s$, $T$ exhibits
sharp peaks or dips near the dual-CIRs due to comparable contributions of both s-
and p-waves. In contrast to the single-mode regime there is no
a complete transmission at the dual CIR points. Our results for
the transmission coefficient versus energy show that in the limit of zero
longitudinal relative energy the distinguishable atoms
behave like bosons, while in the vicinity of the position of a
shape-resonance, they behave like fermions. For larger energies $\varepsilon >> 1$
$T$ is close to unity. Finally we have analyzed the transition
probability as a function of $-V_0$. For values of $V_0$ where $a_p$ is
negligible compared to $a_s$ we find an excellent agreement with the
results for bosonic collisions.  In regions where these two
scattering parameters are comparable, we observe, similar to the fermionic case, sharp
downward- (for elastic collision) and upward- (for inelastic
collision) peaks but with different values.
We conclude with the general statement that our multi-channel scattering results in waveguides
are of immediate relevance to cold or ultracold atomic collisions in atomic waveguides
or impurity scattering in quantum wires. 

\section{Acknowledgments}

S.S. acknowledges a scholarship by the Ministry of Science,
Research and Technology of Iran. V.S.M. acknowledges financial
support by the Landesstiftung Baden-W\"urttemberg in the framework of a 
guest program. Financial support by the Heisenberg-Landau Program is also acknowledged.
P.S. thanks the Deutsche Forschungsgemeinschaft for financial support.

\section{Appendix A. Boundary conditions}

In the limit $r$ tends to zero, $\mathbf{u}$ goes to zero.  This boundary condition can be satisfied easily by putting
$\mathbf{u}(r=0)=\mathbf{0}$ and $\mathbf{u}_{-j}=\mathbf{u}_j$ in the vicinity of $r=0$.  The latter is needed for
approximating the derivatives $\frac{d^2}{dx^2}u(r(x))$ and $\frac{d}{dx}u(r(x))$ near the point $r=0$.

The boundary condition for $u$ approximating the asymptotic form (\ref{full asymptotic wave}) at large $r$ can
be written in the form
\begin{equation}\label{right-side boundary condition}
\mathbf{u}_j+\alpha_j^{(1)}\mathbf{u}_{j-1}+\alpha_j^{(2)}\mathbf{u}_{j-2}+
\alpha_j^{(3)}\mathbf{u}_{j-3}+\alpha_j^{(4)}\mathbf{u}_{j-4}=\mathbf{g}_j\quad\quad j=N-2, N-1, N
\end{equation}
where $\alpha_j$s are diagonal $N_{\theta}\times N_{\theta}$ matrices and $\mathbf{g}_j$ is a $N_\theta$-dimensional vector.
The above boundary-conditions are constructed by eliminating the unknown amplitudes $f_{nn'}$ from the asymptotic
equations (\ref{full asymptotic wave}) written  for a few
$r_j$s neighbouring to the point $r_N=R$.  This gives the values for the coefficients $\alpha_j$s and $\mathbf{g}_j$.
In general for up to four open channels we have
\begin{equation}
\left [ \alpha_j^{(l)}\right ]_{mm'}=(-1)^i\left ( T_j^{l-1,m}+T_j^{l,m}\right )\frac{r_j}{r_{j-l}}\frac{\phi_{n_{e}}(\rho^m_{j})}{\phi_{n_{e}}(\rho^m_{j-l})}e^{ik_{n_{e}}(|z^m_j|-|z^m_{j-l}|)}\delta_{mm'}, \quad\quad l=1,2,3,4
\end{equation}
and
\begin{equation}\label{g}
\left [ \mathbf{g}_j\right ]_{m}=2\sqrt{\lambda_m}r_j\phi_{n_{e}}(\rho^m_{j})e^{ik_{n_{e}}|z^m_j|}\sum_{n=0}^{n_e}\sum_{l=0}^4(-1)^ie^{-ik_{n_{e}}|z^m_{j-l}|}e^{ik_nz^m_{j-l}}
(T_j^{l-1,m}+T_j^{l,m})\frac{\phi_n(\rho^m_{j-l})}{\phi_{n_{e}}(\rho^m_{j-l})}
\end{equation}
For the case of bosonic (fermionic) collisions we must consider just the even (odd) part of the equation (\ref{g}), i.e.
we just need to replace the term $e^{ik_nz^m_{j-l}}$ by $\cos k_nz^m_{j-l}$ ($i\sin k_nz^m_{j-l}$).
Here $\phi_{n}(\rho)=\phi_{n,0}(\rho,\varphi)$ is the eigenfunction of the transverse trapping
Hamiltonian (see Eq.(\ref{transvers Hamil. eigenfunction})), $\rho^m_j=r_j\sin \theta_m$, $z^m_j=r_j\cos \theta_m$, $n$ is
the channel number of the initial state, and $n_e$ is the number of transversely excited open channels.
The complex numbers $T^{i,m}_j$s are given through
\begin{equation}
T_j^{l,m}=\delta_{l,0}+\delta_{l,1}(a^m_j+b^m_j+c^m_j)+\delta_{l,2}(a^m_jb^m_{j-1}+a^m_jc^m_{j-1}+b^m_jc^m_{j-1})+\delta_{l,3}a^m_jb^m_{j-1}c^m_{j-2} \end{equation}
If $n_{e}<1$ (a single-mode regime) $a^m_j=0$, otherwise
\begin{eqnarray}
a^m_j=\left \{\Xi^m_{0,j}-(1+b^m_j+c^m_j)\Xi^m_{0,j-1}+(b^m_j+c^m_j+b^m_jc^m_{j-1})\Xi^m_{0,j-2}-b^m_jc^m_{j-1}\Xi^m_{0,j-3}\right \}\times\nonumber\\\left \{\Xi^m_{0,j-1}-(1+b^m_j+c^m_j)\Xi^m_{0,j-2}+(b^m_j+c^m_j+b^m_jc^m_{j-1})\Xi^m_{0,j-3}-b^m_jc^m_{j-1}\Xi^m_{0,j-4}\right \}^{-1}
\end{eqnarray}
if $n_{e}<2$, $b^m_j=0$, otherwise
\begin{equation}
b^m_j=\left \{\Xi^m_{1,j}-(1+c^m_j)\Xi^m_{1,j-1}+c^m_j\Xi^m_{1,j-2}\right \}\times\left \{\Xi^m_{1,j-1}-(1+c^m_j)\Xi^m_{1,j-2}+c^m_j\Xi^m_{1,j-3}\right \}^{-1}
\end{equation}
if $n_{e}<3$, $c^m_j=0$, otherwise
\begin{equation}
c^m_j=\left \{\Xi^m_{2,j}-\Xi^m_{2,j-1}\right \}\times\left \{\Xi^m_{2,j-1}-\Xi^m_{2,j-2}\right \}^{-1}
\end{equation}
there $\Xi^m_{n,j}=e^{i(k_n-k_{n_e})|z^m_j|}\frac{\phi_n(\rho^m_j)}{\phi_{n_e}(\rho^m_j)}$.  $k_n$ and $k_{n_e}$ are given by eq.(\ref{wave vector k}).

\section{Appendix B. Fast implicit matrix algorithm}

Following the idea of the LU-decomposition \cite{Press} and the sweep method \cite{Gelfand} (or
the Thomas algorithm \cite{fia}), we search the solution of the system of $N$ vector equations with any coefficient being
a $N_{\theta}\times N_{\theta}$  matrix (\ref{LU}) in the form
\begin{equation}\label{vector solution}
\mathbf{u}_j=C^{(1)}_j\mathbf{u}_{j+1}+C^{(2)}_j\mathbf{u}_{j+2}+C^{(3)}_j\mathbf{u}_{j+3},\quad\quad j=1,...,N-3
\end{equation}
Here we define $\mathbf{u}(r(x_j))$ as $\mathbf{u}_j$ for simplicity.  The $C_j$s are unknown
$N_{\theta}\times N_{\theta}$  matrices.  To find the solution, first we should calculate the unknown $C_j$ matrices.
The plan is the following: Due to eq.(\ref{vector solution}) we have
\begin{equation}
\mathbf{u}_{j-p}=C^{(1)}_{j-p}\mathbf{u}_{j-p+1}+C^{(2)}_{j-p}\mathbf{u}_{j-p+2}+C^{(3)}_{j-p}\mathbf{u}_{j-p+3},\quad\quad p=1, 2,
3\,\,,
\end{equation}
then one obtains
\begin{equation}\label{uj-1}
\mathbf{u}_{j-1}=C^{(1)}_{j-1}\mathbf{u}_{j}+C^{(2)}_{j-1}\mathbf{u}_{j+1}+C^{(3)}_{j-1}\mathbf{u}_{j+2}
\end{equation}
\begin{equation}\label{uj-2}
\mathbf{u}_{j-2}=\left [C^{(1)}_{j-2}C^{(1)}_{j-1}+C^{(2)}_{j-2}\right ]\mathbf{u}_{j}+\left [C^{(1)}_{j-2}C^{(2)}_{j-1}+C^{(3)}_{j-2}\right ]\mathbf{u}_{j+1}
+C^{(1)}_{j-2}C^{(3)}_{j-1}\mathbf{u}_{j+2}
\end{equation}
and
\begin{eqnarray}\label{uj-3}
\mathbf{u}_{j-3}=\left [C^{(1)}_{j-3}C^{(1)}_{j-2}C^{(1)}_{j-1}+C^{(1)}_{j-3}C^{(2)}_{j-2}+C^{(2)}_{j-2}C^{(1)}_{j-1}+C^{(3)}_{j-2}\right ]\mathbf{u}_{j}\nonumber\\ {}+\left [C^{(1)}_{j-3}C^{(1)}_{j-2}C^{(2)}_{j-1}+C^{(1)}_{j-3}C^{(3)}_{j-2}+C^{(2)}_{j-2}C^{(2)}_{j-1}\right ]\mathbf{u}_{j+1}
\nonumber\\ {}+\left [C^{(1)}_{j-3}C^{(1)}_{j-2}C^{(3)}_{j-1}+C^{(2)}_{j-2}C^{(3)}_{j-1}\right ]\mathbf{u}_{j+2}
\end{eqnarray}
By substituting $u_j$ defined by Eqs.(\ref{uj-1}-\ref{uj-3}) into Eq.(\ref{LU}) one can calculate $\mathbf{u}_j$ in terms of
$\mathbf{u}_{j+1}$, $\mathbf{u}_{j+2}$ and $\mathbf{u}_{j+3}$.  Then, by comparing with Eq.(\ref{vector solution}) we
find a recurrence formula for calculating the unknown matrices $C_j$:
\begin{eqnarray}\label{C1}
-D C_j^{(1)}= \mathbb{A}^j_{j-3}\left [C^{(1)}_{j-3}C^{(1)}_{j-2}C^{(2)}_{j-1}+C^{(1)}_{j-3}C^{(3)}_{j-2}+C^{(2)}_{j-3}C^{(2)}_{j-1}\right ]\nonumber\\ {}+\mathbb{A}^j_{j-2}\left [C^{(1)}_{j-2}C^{(2)}_{j-1}+C^{(3)}_{j-2}\right ] +\mathbb{A}^j_{j-1}C^{(2)}_{j-1}+\mathbb{A}^j_{j+1}
\end{eqnarray}
\begin{eqnarray}\label{C2}
-D C_j^{(2)}= \mathbb{A}^j_{j-3}\left [C^{(1)}_{j-3}C^{(1)}_{j-2}C^{(3)}_{j-1}+C^{(2)}_{j-3}C^{(3)}_{j-1}\right ]\nonumber\\ {}+\mathbb{A}^j_{j-2}C^{(1)}_{j-2}C^{(3)}_{j-1} +\mathbb{A}^j_{j-1}C^{(3)}_{j-1}+\mathbb{A}^j_{j+2}
\end{eqnarray}
and
\begin{eqnarray}\label{C3}
-D C_j^{(3)}= \mathbb{A}^j_{j+3}.
\end{eqnarray}
Here
\begin{eqnarray}\label{C4}
D= \mathbb{A}^j_{j-3}\left [C^{(1)}_{j-3}C^{(1)}_{j-2}C^{(1)}_{j-1}+C^{(1)}_{j-3}C^{(2)}_{j-2}+C^{(2)}_{j-3}C^{(1)}_{j-1}+C^{(3)}_{j-3}\right ]\nonumber\\ {}+\mathbb{A}^j_{j-2}\left [C^{(1)}_{j-2}C^{(1)}_{j-1}+C^{(2)}_{j-2}\right ] +\mathbb{A}^j_{j-1}C^{(1)}_{j-1}+\mathbb{A}^j_{j}+2(\epsilon I -V_j)
\end{eqnarray}
By using the left-side boundary conditions and Eq.(\ref{vector solution}) one can calculate the $C_j$ matrices for $j=1, 2$ and $3$.
Then by using Eqs.(\ref{C4}-\ref{C1}) we calculate all the matrices $C_j$.  Subsequently by using the right-side boundary
conditions (\ref{right-side boundary condition}) and recurrence formula (\ref{vector solution}) we first calculate
$\mathbf{u}_j$ for $j=N-2, N-1$ and $N$ and then $\mathbf{u}_j$ for $j=1, ...,N-3$.


\begin{thebibliography}{40}
\bibitem{traps} R.~Grimm, M.~Weidem\"uller, and Y.B.~Ovchinnikov, Adv. At. Mol. Opt. Phys. {\bf 42}, 95 (2000).
\bibitem{chips1} R.~Folman {\it et al.},Adv. At. Mol. Opt. Phys. {\bf 42}, 95 (2000).
\bibitem{chips2} J.~Reichel, Appl. Phys. B: Laser Opt. {\bf 74}, 469 (2002).
\bibitem{chips3} J.~Fortagh {\it et al.}, Rev. Mod. Phys. {\bf 79}, 235 (2007).
\bibitem{bolda} E.L. Bolda, E. Tiesinga and P.S. Julienne, Phys.Rev.A {\bf{66}}, 013403 (2002).
\bibitem{stock1} R. Stock, I.H. Deutsch and E.L. Bolda, Phys.Rev.Lett.{\bf{91}}, 183201 (2003).
\bibitem{yurovsky1} V.A. Yurovsky, Phys.Rev.A{\bf{71}}, 012709 (2005).
\bibitem{stock2} R. Stock and I.H. Deutsch, Phys.Rev.A {\bf{73}}, 32701 (2006).
\bibitem{yurovsky2} V.A. Yurovsky and Y.H. Band, Phys.Rev.A {\bf{75}}, 012717 (2007).
\bibitem{naidon} P. Naidon {\it{et al}}, New J. Phys. {\bf{9}}, 19 (2007).
\bibitem{bhongale} S.G. Bhongale, S.J.J.M.F. Kokkelmans and I.H. Deutsch, arXiv:0712.2070v1, physics:atom-ph.
\bibitem{yurovsky3} V.A. Yurovsky, M. Ol'shanii and D.S. Weiss, Adv.At.Mol.Opt.Phys. {\bf{55}}, 61 (2007).
\bibitem{Olshanii0} M.~Olshanii, Phys. Rev. Lett. {\bf 81}, 938 (1998).
\bibitem{Olshanii1} T.~Bergeman, M.G.~Moore, and M.~Olshanii, Phys. Rev. Lett. {\bf 91}, 163201 (2003).
\bibitem{Mora1a} C.~Mora, R.~Egger, A.O.~Gogolin, and A.~Komnik, Phys. Rev. Lett. {\bf 93}, 170403 (2004).
\bibitem{Mora1b} C.~Mora, R.~Egger, and A.O.~Gogolin, Phys. Rev. {\bf A71} 052705 (2005).
\bibitem{Mora2} C.~Mora, A.~Komnik, R.~Egger, and A.O.~Gogolin, 2005 Preprint cond-mat/0501641.
\bibitem{Granger} B.E.~Granger and D.~Blume, Phys.Rev. Lett. {\bf 92}, 133202 (2004).
\bibitem{Kinoshita} T.~Kinoshita, T.~Wenger and D.S.~Weiss, Science {\bf 305}, 1125 (2004).
\bibitem{Paredes} B.~Paredes et al, Nature {\bf 429}, 277 (2004).
\bibitem{Guenter} K.~G\"{u}nter et al, Phys. Rev. Lett. {\bf 95}, 230401 (2005).
\bibitem{Kim1} J.I.~Kim, J.~Schmiedmayer, and P.~Schmelcher, Phys. Rev. {\bf A72}, 042711 (2005).
\bibitem{Egger} V.~Peano, M.~Thorwart, C.~Mora, and R.~Egger, New J. Phys. {\bf 7}, 1 (2005).
\bibitem{Kim2} J.I.~Kim, V.S.~Melezhik, and P.~Schmelcher, Phys. Rev. Lett. {\bf 97}, 193203 (2006).
\bibitem{Kim3} J.I.~Kim, V.S.~Melezhik, and P.~Schmelcher, Rep. Progr. Theor. Phys. Supp. {\bf 166}, 159 (2007).
\bibitem{Mel2007} V.S.~Melezhik, J.I.~Kim, and P.~Schmelcher, Phys. Rev. {\bf A76}, 053611 (2007).
\bibitem{Lupu1} A.~Lupu-Sax, Quantum Scattering Theory and Applications, PhD Thesis, Harvard University (1998).
\bibitem{Lupu2} M.G.E. da Luz, A.S. Lupu-Sax and E.J. Heller, Phys.Rev.E {\bf{56}}, 2496 (1997)
\bibitem{Olshanii2} M.G.~Moore, T.~Bergeman and M.~Olshanii, J. Phys. IV {\bf 116}, 69 (2004), Lecture courses
Les Houches School on 'Quantum Gases in Low Dimensions' (2003).
\bibitem{Mel91} V.S.~Melezhik, J. Comput. Phys. {\bf 92}, 67 (1991).
\bibitem{Mel2003} V.S.~Melezhik and Chi-Yu Hu, Phys. Rev. Lett. {\bf 90}, 083202 (2003).
\bibitem{Mel97} V.S.~Melezhik, Phys. Lett. {\bf A230}, 203 (1997).
\bibitem{Press}W.H.~Press, S.A.~Teukolsky, W.T.~Vetterling, and B.P.~Flannery, {\it Numerical Recipes}, Cambridge University Press, 1992.
\bibitem{Gelfand} I.M.~Gelfand and S.V.~Fomin,{\it Calculus of Variations}, Dover Publications, 2000.
\bibitem{fia} G.N.~Bruse et al., Petrol. Trans. AIME {\bf 198}, 79 (1953).

\end{thebibliography}
\end{document}